\documentclass[aps,reprint,prb,floatfix,footinbib]{revtex4-2}
\usepackage{hyperref}
\hypersetup{
  breaklinks=true, 
  colorlinks=true, 
  pdfusetitle=true, 
}
\usepackage{float}
\usepackage{epsfig}
\usepackage{graphicx}
\usepackage{amsmath}
\usepackage{amsfonts}
\usepackage{amssymb}
\usepackage{bm}
\usepackage{bbold}
\usepackage{epsfig}
\usepackage{graphicx}
\usepackage{color}
\usepackage{pictex}
\usepackage{mathtools}
\usepackage{extarrows}
\usepackage{footnote}
\usepackage{footmisc}
\newcommand{\s}{\scriptscriptstyle}
\begin{document}
\title{Brane order and quantum magnetism in modulated anisotropic ladders}
\author{Toplal Pandey}
\affiliation{Department of Physics, Laurentian University, Sudbury, Ontario,
P3E 2C6 Canada}
\author{Gennady Y. Chitov}
\affiliation{Institut quantique \& D\'{e}partement de physique, Universit\'{e} de Sherbrooke,
Sherbrooke, Qu\'{e}bec, J1K 2R1 Canada}
\date{\today}

%
%
\begin{abstract}
Two-leg spin-$\frac12$ ladders with anisotropy and two different dimerization patterns are analyzed at zero temperature. This model is equivalent to a modulated interacting (Kitaev) ladder.  The Hartree-Fock mean-field approximation reduces the model to a sum of two quadratic effective Majorana Hamiltonians, which are dual to two quantum transverse XY chains.
The mapping between the effective Hamiltonian of the ladder and a pair of chains considerably simplifies calculations the order parameters and analysis of the hidden symmetry breaking.  The ground-state phase diagram of the staggered ladder contains nine phases,
four of them are conventional antiferromagnets, while the other five possess non-local brane orders. Using the dualities and the newly found exact results for the local and string order parameters of the transverse XY chains, we were able to find analytically all the magnetizations and
the brane order parameters for the staggered case, as functions of the renormalized couplings of the effective Hamiltonian.
The columnar ladder has three ground-state phases and does not possess magnetic long-ranged order. The brane order parameters for these three phases are calculated numerically from the Toeplitz determinants. All brane-ordered phases are spin liquids with identified distinct order parameters, winding numbers, and sets of the Majorana edge modes. Disorder lines and the special points of disentanglement are found for both dimerization patterns. We expect this study to motivate the search for the real spin-Peierls anisotropic ladder compounds which manifest predicted properties.
\end{abstract}
\maketitle

%
%
%
\section{Introduction}\label{Intro}
%
%
%
%

Spin ladders have been the focus of significant theoretical interest
in the past several decades \cite{Dagotto:1999,*Dagotto:1996,Giamarchi:2004}.
One of the most peculiar properties of spin ladders is that the existence of a gap
(i.e. mass) depends on the number of legs. The spin excitations in an isotropic (XXX)
antiferromagnetic $m$-leg spin-$\frac12$ ladder are gapped if $m$ is even, and the system is
gapless (quantum critical) when the number of legs $m$ is odd. The
even-$m$-leg ladders are interesting examples of spin liquids,
where a gap is not accompanied by a local long-ranged order or apparent
symmetry breaking.

The purpose of this study is to explore the ground-state phase diagram and the nature of the ``hidden'' (non-local)
orders in the massive phases of the two-leg antiferromagnetic Heisenberg spin-$\frac12$ ladders whose isotropic
Hamiltonian is perturbed by the relevant terms as dimerization and anisotropy (XXX $\mapsto$ XYZ).
It is known even for a single gapless XXX chain that the interplay of dimerization and anisotropy, each of which
leads to a gap opening, can result in gaplessness or quantum criticality in some range of parameters \cite{Takahashi:1999}.
Similarly, the gapped isotropic two-leg ladder perturbed by dimerization and anisotropy, can demonstrate various
phases and quantum phase transitions in the space of relevant parameters.

When two chains are coupled into a ladder, the system is gapped due to the relevant interchain exchange coupling \cite{Giamarchi:2004}.
Subtle interplay of relevant terms may result in that the dimerized two- (or three-) leg ladders made out of gapped spin chains, can be gapless.
The criticality (gaplessness) in the dimerized two- and three-leg ladders was first conjectured in \cite{Delgado:1996} and has been confirmed by
subsequent numerical and analytical work
\cite{Delgado:1998,Kotov:1999,Cabra:1999,Nersesyan:2000,Okamoto:2003,Nakamura:2003,Delgado:2007,*Delgado:2008,*Delgado:2008JPA,Chitov:2008,Chitov:2011PRB,
Chitov:2017JSM,Xu:2012,Ding:2020,Fu:2021}.
The main interest in dimerized ladders comes from the real experiments on the spin-Peierls ladder-type compounds, see, e.g.,  \cite{Mayaffre:2000},
so the spin-Peierls transitions were analyzed in the ladder models, see, e.g., \cite{Chitov:2007,Ding:2009,Xu:2012}.

A drawback for the experimental observations of the predicted quantum phase transitions in the dimerized ladder is that if the said dimerization occurs due to the spin-Peierls transition, and it is not a built-in property of the Hamiltonian, then the ladder locks itself into the plainly gapped columnar dimerization pattern which has lower energy \cite{Chitov:2008}, rather then into the energetically unfavorable staggered pattern (see Fig.~\ref{DimLad}) which can demonstrate the quantum criticality predicted in \cite{Delgado:1996}. As we show in the present work, in a more general ladder model with the spin exchange $xy$-anisotropy, the quantum phase transitions occur for the both types of the dimerization patterns, which  opens the possibility for experimental observations of the predicted phases.

On the theoretical side, we are not aware of similar studies of the spin ladders with the anisotropy and dimerization, however there has been work done on the ground-state phases and entanglement in the Kitaev fermionic ladders with modulations and anomalous terms \cite{ZhouShen:2011,Zhou:2016,Wang:2016,Hung:2017,Zhou:2017,Citro:2018,Nersesyan:2020,Nehra:2018,Nehra:2020}. Such fermionic models are the closest counteparts of the spin ladders in question, since the models with $\frac12$-spins or spinless fermions can be mapped onto each other by a judiciously chosen Jordan-Wigner transformation. Whether one deals with the spin or fermionic ladder, the fundamental question to answer is the nature of the phases of the model, that is, the order corresponding to a given phase.

The key notions of the Landau paradigm are the order parameter and the symmetry it breaks
spontaneously  \cite{LandauV5}. There has been a huge recent effort to understand whether
various quantum spin liquids, frustrated magnets, topological and Mott insulators, etc \cite{Fradkin:2013,Asboth:2016,Shen:2017,RyuSchnyder:2010,Montorsi:2012}, which often lack conventional local order even at zero temperature, can be dealt within the Landau framework or some new
paradigms are needed  \cite{Wen:2017}.

The line pursued in the present study is that the extended Landau theory which incorporates the notions of nonlocal (string) order \cite{denNijs:1989} and spontaneous breaking of hidden symmetry \cite{Oshikawa:1992,*Kennedy:1992,*Kohmoto:1992}, remains instrumental even for nonconventional orders \cite{Chitov:2017JSM,Chitov:2018,Chitov:2019,Chitov:2020}.
The local and nonlocal string order parameters in the extended formalism are related by duality, and
probing a phase transition and related emerging order is a problem of appropriate choice of variables
\cite{ChenHu:2007,Xiang:2007,NussinovChen:2008,Nussinov:2009,*Nussinov:2013,Smacchia:2011,Chitov:2017JSM,Chitov:2018,Chitov:2019,Chitov:2020}.

The latter point becomes painfully obvious if we take a paradigmatic toy model much discussed in the recent literature, namely the so-called Kitaev fermionic chain \cite{Kitaev:2001}.  It has two phases with all attributes of the topological order: they are gapped, degenerate, no apparent local order or symmetry breaking, non-trivial topological winding numbers $N_w = \pm 1$, and even the zero-energy Majorana edge modes. However, as emphasized by Fendley \cite{Fendley:2012}, with a flip of a coin the physics may be rendered plain-vanilla-like. The fermionic model in the spin representation is the well-known $XY$ chain in transverse field with two-fold degenerate (anti)ferromagnetic phases, clean cut symmetry breaking,  and the conventional exactly-known local order parameters $m_{x,y}$ (see e.g., \cite{Franchini:2017}). Even the Majorana modes, if there is interest, can be recovered. They resurge in the spin framework as the surface (edge) magnetizations \cite{Peschel:1984,Karevski:2000}.

The very important point is that the string order yields a proper order parameter in the
sense of Landau: its critical index $\beta$ along with other critical indices satisfy
the standard (hyper)scaling relations. This is quite simple to establish for those quantum models which are equivalent to the one-dimensional free fermions, that is to the 2D Ising universality class. There is a rare example of analytical results available for an interacting model: from the bosonization calculations for the string parameters of the dimerized XXX chain, due to Hida \cite{Hida:1992}, one can find the critical indices $\beta=1/12$, $\eta=1/4$, and  $\nu=2/3$,  and verify that they
correspond to a special parametric point of the eight-vertex model \cite{Baxter:2007} and satisfy all scaling relations. The string order parameters can be used
to study spins, fermions, and bosons \cite{Berg:2008}, and even be observed \cite{Endres:2011}.

The notion of the string order parameter initially defined for a chain  \cite{denNijs:1989}, was later generalized for spin ladders
\cite{Nishiyama:1995,Watanabe:1995,Shelton:1996,White:1996,Kim:2000,*Fath:2001,*Kim:2008,Nakamura:2003b}. However to systematically probe non-local order beyond 1D, one needs to define the brane order parameters \cite{NussinovChen:2008,Cirac:2008,Rath:2013,Bahri:2014,Montorsi:2016,*Montorsi:2017,*Montorsi:2019}, and that is the concept we use in this work. As shown below in case of the two-leg ladders, the brane parameters probe genuine non-local order both along the chains and the rungs.
\footnote{\label{SOPBOP} In the connection with the earlier related work on the ladders \cite{Chitov:2011PRB,Chitov:2017JSM}, the string order parameters defined there for the ladders are just special cases of the branes considered in the present study. However the current use of the term ``brane'' is more accurate and appropriate.}

The spin ladder is equivalent to an interacting fermionic model which is treated in this work within a Hartree-Fock mean-field approach. Such mean-field theory is
known to be quite accurate even quantitatively for spin ladders \cite{Azzouz:1993,*Azzouz:1994,Dai:1998,Hori:2004,Nunner:2004}, and was previously successfully applied
to study the quantum phase transitions with non-local orders in the dimerized ladders \cite{Chitov:2007,Chitov:2008,Chitov:2017JSM,Chen:2012,Li:2013}. The key goal of this approximation is to obtain an effective quadratic fermionic Hamiltonian. Its spectrum yields the ground-state phase diagram, topological winding numbers. This Hamiltonian  is used to calculate the thermodynamic quantities, and in particular, the order parameters. The progress in calculation of the brane order parameters is made in the current work by using the duality transformations and mapping the effective fermionic Hamiltonian of the ladder first onto two decoupled Majorana Hamiltonians, and then onto two decoupled modulated XY chains. As a result, the brane order parameters of the ladder are expressed via the local or string order parameters of the chains, and found analytically in a closed form for the ladder with staggered dimerization. For the columnar dimerized ladder those parameters are calculated numerically from the Toeplitz determinants.

The rest of the paper is organized as follows:  In Sec.~\ref{Model} the spin ladder model is defined, and the effective fermionic Hamiltonian is introduced along with its mean-field (renormalized) parameters. (The derivation of the effective Hamiltonian and details on the mean-field equations are presented in
Appendix~\ref{AppA}). Sections  \ref{SpecMod} and \ref{OPsAll} contain most of the formalism and results. The physical quantities which do not belong to the
standard set of parameters of the Landau framework: topological winding numbers, zero-energy Majorana edge states, entanglement, are presented in Sec.~\ref{TopNum}. Appendix~\ref{AppB} contains several new exact results for the order parameters in the XY chain with transverse fields, applied to calculate the brane orders.  The results are summarized in the concluding Sec.~\ref{Concl}.

%
%
%
%
\section{Model and the effective mean-field Hamiltonian}\label{Model}
\begin{figure}[h]
\epsfig{file=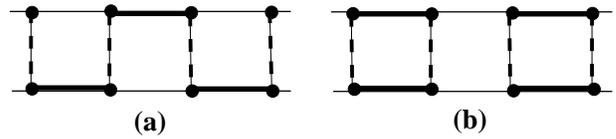,width=0.45\textwidth,angle=0} \caption{Dimerized
two-leg ladder. Bold or thin lines represent the stronger or weaker
chain coupling; dashed lines represent rung coupling $J_\bot$,
respectively.  Dimerization patterns: (a) - staggered; (b)-
columnar.} \label{DimLad}
\end{figure}

%
%
%
In this paper we analyze the Heisenberg spin-$\frac12$ two-leg ladder with intrinsic dimerization and the $xy$ spin anisotropy
at zero temperature. The ladder Hamiltonian is given by:
\begin{eqnarray}
\label{Ham}
H &=& \sum_{n=1}^{N}\sum_{\alpha=1}^{2} \Big\{ J_{\alpha}(n)
\mathbf{S}_{\alpha}(n) \cdot \mathbf{S}_{\alpha}(n+1) \nonumber \\
&+&  J \gamma  \big[ S^x_{\alpha}(n) S^x_{\alpha}(n+1)-  S^y_{\alpha}(n) S^y_{\alpha}(n+1) \big] \Big\} \nonumber \\
&+& J_\bot \sum_{n=1}^{N} \mathbf{S}_{\alpha}(n) \cdot \mathbf{S}_{\alpha+1}(n).
\end{eqnarray}
The dimerization and anisotropy are assumed along the chains only, with the rung coupling $J_\bot$ taken as constant. All the spin exchange couplings
are antiferromagnetic. The two possible dimerization patterns shown in Fig.~\ref{DimLad} are defined as:
\begin{equation}
\label{Couplings}
J_{\alpha}(n)=
\left\{
\begin{array}{lr}
J[1+(-1)^{n+\alpha} \delta],&~~~~~~~\mbox{staggered}\\
J[1+(-1)^{n} \delta].&~~~~~~\mbox{columnar}\\
\end{array}
\right.
\end{equation}
The spin operators $\mathbf{S}= \frac12 \boldsymbol{\sigma}$ are defined in terms of  the standard Pauli matrices $\boldsymbol{\sigma}$ on the chains $(\alpha=1,2)$, the bond alternation parameter $|\delta|\leq1$, and $\gamma$ is the anisotropy parameter.
%
%

The spin ladder will be treated within a mean-field approach  \cite{Azzouz:1993,*Azzouz:1994} which consists of two steps:
first, one maps the spin Hamiltonian onto an interacting model of spinless fermions via a Jordan-Wigner transformation; second,
the interaction terms are decoupled via the Hartree-Fock approximation. This mean-field approach is confirmed
to be qualitatively and even quantitatively adequate for analysis of similar ladder models \cite{Chitov:2007,Chitov:2008,Chitov:2017JSM}.
Relegating the technical details to Appendix~\ref{AppA}, the spin model is reduced to  the quadratic
effective fermionic Hamiltonian
\begin{widetext}
\begin{equation}
\label{Ham2L3}
 H_{\s MF}=\frac12 \sum_{n} \Big\{  \sum_\alpha (-1)^{n+\alpha-1}\big[ J_{\alpha \s R}(n)c^{\dag}_{\alpha}(n) c_{\alpha}(n+1)+\Gamma_{ \s R}(n)c^{\dag}_{\alpha}(n)c^{\dag}_{\alpha}(n+1) \big]
+J_{\perp \s R}(n) c^{\dag}_{1}(n)c_{2}(n)+h.c. \Big\} +2N \mathcal{C}~.
\end{equation}
\end{widetext}
We have introduced the renormalized couplings:
\begin{equation}
J_{\alpha \s R}(n)=
\left\{
\begin{array}{lr}
J[t_{\s R}+(-1)^{n+\alpha}\delta_{\s R}],&~~~~~~~\mbox{staggered}\\
J[t_{\s R}+(-1)^{n}\delta_{\s R}],&~~~~~~\mbox{columnar}\\
\end{array}
\right.
\end{equation}
\begin{equation}
\Gamma_{\alpha \s R}(n)=
\left\{
\begin{array}{lr}
J[\gamma_{\s R}+(-1)^{n+\alpha}\gamma_{a \s R}],&~~~~~~~\mbox{staggered}\\
J[\gamma_{\s R}+(-1)^{n}\gamma_{a\s R}],&~~~~~~\mbox{columnar}\\
\end{array}
\right.
\end{equation}
and
\begin{equation}
\label{JperpR}
J_{\perp \s R}(n)=J_{\perp}(1+2t_{\perp})~,
\end{equation}
along with the renormalized model's parameters:
\begin{eqnarray}
  \label{tR}
  t_{\s R} &=& 1+2(\mathcal{K}+\delta^2\eta),\\
  \delta_{\s R} &=& \delta \big(1+2(\mathcal{K}+\eta)\big),\\
  \label{dR}
  \gamma_{\s R} &=& \gamma-2(P-\delta^2\eta_{p}),\\
  \label{rR}
  \gamma_{a\s R} &=& -2 \delta(P-\eta_{p}) ~.
  \label{raR}
\end{eqnarray}
The constant term is
 \begin{equation}
 \mathcal{C}= \mathcal{K}^2-P^2+\delta^2(\eta^2-\eta_{p}^2)+2\delta^2(\mathcal{K} \eta+
P\eta_{p})+\frac{1}{2}J_{\perp}t_{\perp}^2~.
 \end{equation}
The definitions and the self-consistent equations for the mean-field parameters entering above relations are given in Appendix~\ref{AppA}.
There we also present formulas and numerical results for the mean-field parameters and renormalized couplings and their relations to the bare parameters of the
microscopic Hamiltonian. In the following we will be working with the effective Hamiltonian (\ref{Ham2L3}), and for notational simplicity
we drop the subscript $R$ in its parameters from now on, keeping in mind that we deal with the renormalized parameters
\eqref{tR}-\eqref{raR}. We also express the dimensional quantities in the units of $J$ from now.

%
%
\begin{figure}[h]
\includegraphics[width=6.0 cm]{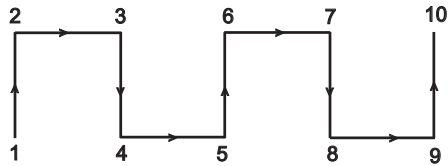}
\caption{The path of Jordan-Wigner transformation used for fermionization of the two-leg ladder. This is also the path used to map the sites of the two-leg ladder onto a snake-like chain.}
\label{SnakePath}
\end{figure}
%
%

Note that (\ref{Ham2L3}) can be also viewed as a tight-binding Hamiltonian of the Kitaev-Majorana ladder \cite{Citro:2018}.
To get rid of the factor $(-1)^n$ in front of the quadratic terms of the Hamiltonian (\ref{Ham2L3}) we use the canonical transformation
\begin{equation}
\label{phiN}
  c_{\alpha}(n) \mapsto e^{i \phi_n} c_{\alpha}(n), ~\mathrm{where}~ \phi_{n+1}=\phi_n +\pi n~(\phi_1=0)~.
\end{equation}
We then introduce an extra label to distinguish two fermion species residing on  even/odd sites as
$c_{\alpha}(n) \mapsto c_{\alpha, e/o}$ for $n=2l$ or $n=2l-1$, respectively. Then the Fourier transform of  the Hamiltonian (\ref{Ham2L3})
can be written as
\begin{equation}
H=2NC+\frac12 \sum_{k}\Psi^{\dag}_{k}\mathcal{H}(k)\Psi_{k}~,
\label{single Ham11}
\end{equation}
where the even and odd fermions are unified in the spinor
\begin{widetext}
\begin{equation}
\Psi^{\dag}_{k}=\Big(c^{\dag}_{1,e}(k),c^{\dag}_{1,o}(k),
c^{\dag}_{2,e}(k),c^{\dag}_{2,o}(k)
,c_{1,e}(-k),c_{1,o}(-k),c_{2,e}(-k),c_{2,o}(-k)\Big)
\label{spinor1}
\end{equation}
\end{widetext}
with the wave numbers restricted to the reduced Brillouin zone $k\in[-\pi/2,\pi/2]$, and we set the lattice spacing $a=1$. The $8\times 8$ Hamiltonian matrix $\mathcal{H}(k)$  reads
\begin{equation}
\mathcal{H}(k)=\begin{pmatrix}
\hat{\mathbb{A}}&\hat{\mathbb{B}}\\
\hat{\mathbb{B}}^{\dag}&-\hat{\mathbb{A}}\\
\end{pmatrix}
\label{HAB}
\end{equation}
The explicit form of the $4\times 4$ matrix $\hat{\mathbb{A}}$ depends on the dimerization pattern. For the staggered case
\begin{equation}
\hat{\mathbb{A}}_s=\begin{pmatrix}
\hat{U}^{\s T}& \frac12 J_\perp \mathbb{1} \\
\frac12 J_\perp \mathbb{1}   &-\hat{U}\\
\end{pmatrix}
\label{Ast}
\end{equation}
where
\begin{equation}
\hat{U} \equiv \begin{pmatrix}
0 & t\cos k+i\delta\sin k \\
 t\cos k-i\delta\sin k  &0\\
\end{pmatrix}
\label{Uhat}
\end{equation}
For the columnar pattern
\begin{equation}
\hat{\mathbb{A}}_c=\begin{pmatrix}
\hat{U}& \frac12 J_\perp \mathbb{1} \\
\frac12 J_\perp \mathbb{1}   &-\hat{U}\\
\end{pmatrix}
\label{Ac}
\end{equation}
The $4\times 4$ matrix $\hat{\mathbb{B}}$ is the same for both dimerization patterns and reads
\begin{equation}
\hat{\mathbb{B}}=\begin{pmatrix}
\hat{V}& 0 \\
0  &-\hat{V}^{\s T}\\
\end{pmatrix}
\label{B}
\end{equation}
where
\begin{equation}
\hat{V} \equiv \begin{pmatrix}
0 & \gamma_a \cos k-i\gamma \sin k \\
 -\gamma_a \cos k-i\gamma \sin k  &0\\
\end{pmatrix}
\label{Vhat}
\end{equation}
The staggered anisotropy $\gamma_a$ is absent in the bare Hamiltonian, but could be induced by the mean-field equations.
The actual values are very small in the interesting range of parameters, so $\gamma_a$ will be discarded in the final results.

%
%
%
\section{Spectra, phase diagram, and dual models}\label{SpecMod}
%
%
%
%
\subsection{Staggered ladder}
\begin{figure}[h]
\includegraphics[width=9.0cm]{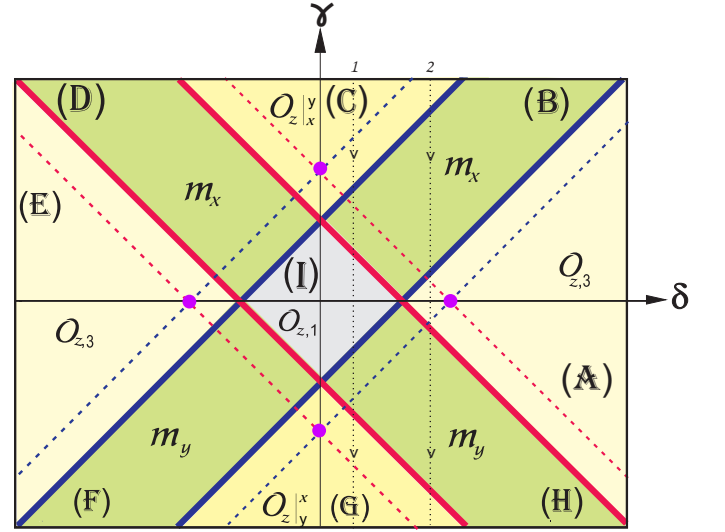}
\caption{Phase diagram of the anisotropic staggered two-leg ladder. Non-vanishing brane and local order parameters are shown in nine regions (A-I) of the ($\delta,\gamma$) parametric plane. The solid bold blue/red lines $(\gamma \mp \delta)^2=J_\perp^2/4$ are the lines of quantum phase transitions (phase boundaries). The dashed blue/red lines denote the even/odd disorder lines $(\gamma \mp \delta)^2=t^2+J_\perp^2/4$ which are bounds of the IC modulations in the even/odd sectors of the Hamiltonian, respectively. The bold magenta dots are the points of disentanglement with the factorized ground state of the Hamiltonian.}
\label{PhaseDiag}
\end{figure}

%
%
%
The $8\times 8$ Hamiltonian matrix \eqref{HAB} (see Eqs.~\eqref{Ast} and \eqref{B}) has eight eigenvalues $\pm E_{\pm\pm}(k)$,  where
\begin{equation}
E_{\pm\pm}(k)=\sqrt{(\gamma_a \pm t)^2\cos^2 k+\Big((\gamma\pm\delta)\sin k\pm\frac{J_{\perp}}{2}\Big)^2}~.
\label{spectra_s}
\end{equation}
 From Eq.~\eqref{spectra_s} we infer that the model  is gapped in general,
 however the gap
 \begin{equation}
 \label{GapSt}
 \Delta=\Big|(\gamma\pm\delta)\pm\frac{J_{\perp}}{2}\Big|
 \end{equation}
at the edge of Brillouin zone $k=\pi/2$  vanishes
on the lines of quantum critical transitions shown in Fig.~\ref{PhaseDiag}:
 \begin{equation}
 \gamma= \pm \Big|\delta\pm\frac{J_{\perp}}{2}\Big|~,
 \label{PhBs}
 \end{equation}
when three relevant perturbations $ \delta, \gamma, J_{\perp}$  cancel, rendering the model gapless.

The phase diagram of the isotropic ($\gamma=0$) dimerized ladder  and quantum phase transitions, accompanied by non-local order parameters in this model,
were actively studied in the literature  \cite{Delgado:1996,Delgado:1998,Kotov:1999,Cabra:1999,Nersesyan:2000,Okamoto:2003,Nakamura:2003,Delgado:2007,*Delgado:2008,*Delgado:2008JPA,Chitov:2008,Chitov:2011PRB,
Chitov:2017JSM,Xu:2012}. The anisotropy ($\gamma \neq 0$) renders model's phase diagram richer: phases with non-local (brane) orders are found along
with conventional antiferromagnetic phases. To the best of our knowledge, these results were not reported before.

To understand the nature of different phases of the phase diagram in Fig.~\ref{PhaseDiag}, it is convenient to relabel the ladder fermionic operators
of the effective mean-field Hamiltonian (\ref{Ham2L3}) according to the snake-like path shown in Fig.~\ref{SnakePath}. This yields
\begin{eqnarray}
  &~&H_{\s MF} = \frac12 \sum_{l=1}^{N}\big\{-(t-\delta)c_{\s 2l-1}^{\dag}c_{\s 2l+2}
                +(t+\delta)c_{\s 2l}^{\dag}c_{\s 2l+1}  \nonumber \\
   &+& \gamma(-c_{\s 2l-1}^{\dag} c_{\s 2l+2}^{\dag}+c_{\s 2l}^{\dag}c_{\s 2l+1}^{\dag})
   +J_{\perp}c_{\s 2l-1}^{\dag}c_{\s 2l}\big\} +h.c.
\label{HMFchain}
\end{eqnarray}
In terms of the Majorana operators \footnote{\label{2Maj} To avoid confusion with earlier related work
\cite{Chitov:2017JSM,Chitov:2018,Chitov:2019,Chitov:2020} note a different definition \eqref{Majop} of Majorana operators.
It corresponds to the tilded Majorana operators defined in the Appendix A of \cite{Chitov:2019}.}
\begin{equation}
2c_{ 2n}^{\dag}=a_{ 2n}+i b_{ 2n-1}
\label{Majop}
\end{equation}
Hamiltonian \eqref{HMFchain} maps onto a sum of two decoupled quadratic Majorana Hamiltonians (Kitaev chains) defined on the even and odd sites of the snake-like chain shown in Fig.~\ref{SnakePath}:
\begin{equation}
\label{EplusO}
H_{\s MF}=H_{o}+H_{e}
\end{equation}
where
\begin{eqnarray}
\label{Habo}
  H_o &=& \frac14 \sum_{l=1}^{N} \big\{-(t-\gamma-\delta) b_{\s 2l+1} a_{\s 2l-1} \nonumber \\
  &+& (t+\gamma+\delta)b_{\s 2l-1}a_{\s 2l+1}
+J_{\perp}b_{\s 2l-1}a_{\s 2l-1}\big\} \\
  H_e &=& \frac14 \sum_{l=1}^{N} \big\{-(t+\gamma-\delta) b_{\s 2l-2} a_{\s 2l+2} \nonumber \\
  &+& (t-\gamma+\delta)b_{\s 2l}a_{\s 2l}
+J_\perp b_{\s 2l-2}a_{\s 2l} \big\}
\label{Habe}
\end{eqnarray}
To advance our analysis we make an inverse Jordan-Wigner transformation of the above Hamiltonians from the Majorana operators to
new dual spins represented by the Pauli matrices $\tilde \tau$. Using the path of Fig.~\ref{SnakePath} to relabel the original ladder spins
$\sigma$, the sequence of the transformations from the original spins to Majoranas and then to dual spins reads
\begin{eqnarray}
  \sigma_{n}^{x}  \sigma_{n+1}^{x} &=& i b_{n-1}  a_{n+1}=\tilde  \tau_{n-1}^{x} \tilde \tau_{n+1}^{x}
  \label{XX-ti} \\
  \sigma_{n}^{y} \sigma_{n+1}^{y} &=& i  b_n  a_n =\tilde \tau_{n}^{z}~.
  \label{YY-ti}
\end{eqnarray}
The dual spins $\tilde \tau$ obey the standard algebra of the Pauli operators, and they reside on the sites of the dual lattice, which can be placed between the sites of the original lattice. Then the odd and even Hamiltonians \eqref{Habo} and \eqref{Habe} become
\begin{eqnarray}
  H_o &=& \frac14 \sum_{l=1}^{N} \big\{(t+\gamma+\delta) \tilde \tau_{\s 2l-1}^{x} \tilde \tau_{\s 2l+1}^{x} \nonumber \\
  &-& (t-\gamma-\delta) \tilde \tau_{\s 2l-1}^{y} \tilde \tau_{\s 2l+1}^{y}+J_{\perp} \tilde \tau_{\s 2l-1}^{z} \big\} \\
\label{HaboT}
  H_e &=& \frac14 \sum_{l=1}^{N} \big\{(t+\gamma-\delta)\tilde \tau_{\s 2l-2}^{x} \tilde \tau_{\s 2l}^{z} \tilde \tau_{\s 2l+2}^{x} \nonumber \\
  &+& J_{\perp} \tilde \tau_{\s 2l-2}^{x} \tilde \tau_{\s 2l}^{x}+(t-\gamma+\delta) \tilde \tau_{\s 2l}^{z} \big\}~.
\label{HabeT}
\end{eqnarray}

Making use of the canonical transformation
\begin{eqnarray}
\label{CanX}
\tau_{\s 2l-1}^{x} &=& \tilde{\tau}_{\s 2l-1}^{x}, \\
\tau_{\s 2l-1}^{y}&=&(-1)^{l}\tilde{\tau}_{\s 2l-1}^{y}, \\
\label{CanY}
\tau_{\s 2l-1}^{z}&=&(-1)^{l}\tilde{\tau}_{\s 2l-1}^{z}.
\label{CanZ}
\end{eqnarray}
we bring the odd sector of the effective Hamiltonian to the from of the XY chain in a staggered transverse field:
\begin{eqnarray}
  H_o &=& \frac14 \sum_{l=1}^{N} \big\{(t+\gamma+\delta)  \tau_{\s 2l-1}^{x}  \tau_{\s 2l+1}^{x} \nonumber \\
  &+& (t-\gamma-\delta)  \tau_{\s 2l-1}^{y}  \tau_{\s 2l+1}^{y}+(-1)^l J_{\perp} \tau_{\s 2l-1}^{z} \big\} ~.
\label{HoXY}
\end{eqnarray}
To simplify the even sector of the Hamiltonian we follow the same steps as above, and then perform an additional dual transformation to $\tilde \mu$-spins
defined as:
\begin{eqnarray}
\label{taumuDual1}
  \tilde \tau_{\s 2l-2}^{x} \tilde \tau_{\s 2l}^{x} &=&  \tilde \mu_{\s 2l-2}^z~,  \\
  \tilde \tau_{\s 2l}^z &=&  \tilde \mu_{\s 2l-2}^y \tilde \mu_{\s 2l}^y~,
\label{taumuDual2}
\end{eqnarray}
followed by the canonical transformation analogous to \eqref{CanX}-\eqref{CanZ}. As a result we obtain the even Hamiltonian in a convenient form
of the XY chain:
\begin{eqnarray}
  H_e &=& \frac14 \sum_{l=1}^{N} \big\{(t+\gamma-\delta)  \mu_{\s 2l-2}^{x}  \mu_{\s 2l}^{x} \nonumber \\
  &+& (t-\gamma+\delta)  \mu_{\s 2l-2}^{y}  \mu_{\s 2l}^{y}+(-1)^l J_{\perp} \mu_{\s 2l-2}^{z} \big\} ~.
\label{HeXY}
\end{eqnarray}

\subsection{Columnar ladder}
\begin{figure}[h]
\includegraphics[width=8.5cm]{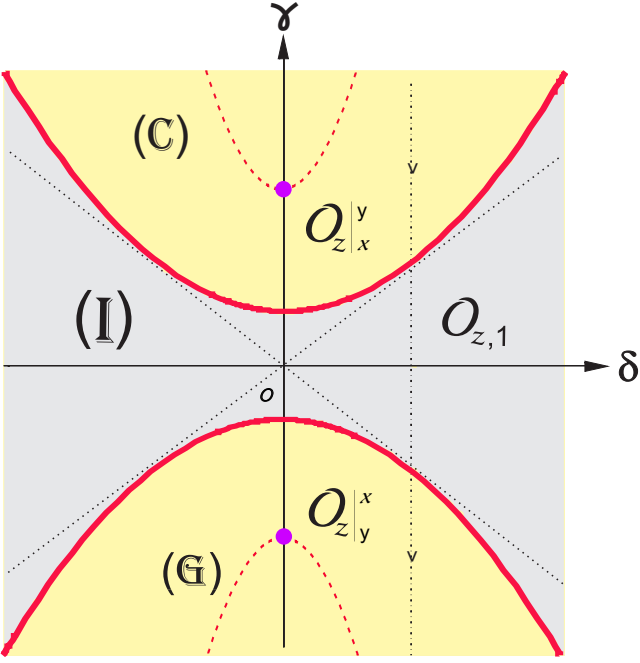}
\caption{Phase diagram of the anisotropic columnar two-leg ladder. Non-vanishing brane order parameters are shown in three regions C, G, and I on the ($\delta,\gamma$) parametric plane. The bold red lines $\gamma^2=\delta^2+J_\perp^2/4$ are the lines of quantum phase transitions.
The dashed red lines denote the disorder lines $ \gamma^2=(1+4t^2/J_\perp^2)(\delta^2+J_\perp^2/4)$ bounding the IC modulations. The bold magenta dots are the points of disentanglement with the factorized ground state of the Hamiltonian.}
\label{Phasecol}
\end{figure}

We diagonalize the $8\times 8$ Hamiltonian  matrix \eqref{HAB}, \eqref{Ac}, \eqref{B} and find four two-fold degenerate
eigenvalues $\pm E_{\pm}(k)$, where
\begin{equation}
E_{\pm}(k)=\sqrt{(\gamma_a \pm t)^2\cos^2 k+\Big(\sqrt{\delta^2\sin^2k+\frac14 J_\perp^2}\pm\gamma\sin k\Big)^2}
\label{spectra_c}
\end{equation}
The spectrum \eqref{spectra_c} has the gap
\begin{equation}
\label{DelCol}
\Delta=\Bigg|\gamma\pm\sqrt{\delta^2+\frac14 J_\perp^2}\Bigg|
\end{equation}
at the edge of the Brillouin zone which vanishes when
\begin{equation}
\gamma^2=\delta^2+\frac14 J_\perp^2~.
\label{ceq_col}
\end{equation}
These curves of quantum criticality (phase boundaries) are plotted in Fig.~\ref{Phasecol}. Anisotropy ($\gamma \neq 0$) brings some new physics contrary to
the ``plain vanilla'' case of the isotropic ($\gamma=0$) columnar-dimerized ladder which is always gapped and locked in the same phase \cite{Chitov:2008,Chitov:2011PRB}.

%
%
%
%

Using relabelling according to the path in Fig.~\ref{SnakePath} and transformations explained in detail in the previous subsection,
the effective mean-field Hamiltonian (\ref{Ham2L3}) for the columnar pattern is mapped onto a sum \eqref{EplusO}
of two decoupled even and odd quadratic Majorana Hamiltonians (Kitaev chains) with
\begin{eqnarray}
  H_o &=& \frac14 \sum_{l=1}^{N} \big\{(t-(-1)^l \delta -\gamma) b_{\s 2l+1} a_{\s 2l-1} \nonumber \\
  &+& (t-(-1)^l \delta +\gamma)b_{\s 2l-1}a_{\s 2l+1}
+J_{\perp}b_{\s 2l-1}a_{\s 2l-1}\big\} \\
\label{HaboC}
  H_e &=& \frac14 \sum_{l=1}^{N} \big\{(t-(-1)^l \delta +\gamma ) b_{\s 2l-2} a_{\s 2l+2} \nonumber \\
  &+& (t-(-1)^l \delta -\gamma)b_{\s 2l}a_{\s 2l}
+J_\perp b_{\s 2l-2}a_{\s 2l} \big\}~.
\label{HabeC}
\end{eqnarray}
In their turn, the Majorana Hamiltonians can be transformed to the dual spins, as defined in the previous subsection, yielding
two equivalent decoupled dimerized XY chains in the staggered transverse fields:
\begin{eqnarray}
\label{HoXYCol}
  H_o &=& \frac14 \sum_{l=1}^{N} \big\{(t-(-1)^l \delta+\gamma)  \tau_{\s 2l-1}^{x}  \tau_{\s 2l+1}^{x} \nonumber \\
  &+& (t-(-1)^l \delta-\gamma)  \tau_{\s 2l-1}^{y}  \tau_{\s 2l+1}^{y}+(-1)^l J_{\perp} \tau_{\s 2l-1}^{z} \big\} \\
  H_e &=& \frac14 \sum_{l=1}^{N} \big\{(t-(-1)^l \delta+\gamma)  \mu_{\s 2l-2}^{x}  \mu_{\s 2l}^{x} \nonumber \\
  &+& (t-(-1)^l \delta-\gamma)  \mu_{\s 2l-2}^{y}  \mu_{\s 2l}^{y}+(-1)^l J_{\perp} \mu_{\s 2l-2}^{z} \big\} .
\label{HeXYCol}
\end{eqnarray}

%
%
%
\section{Phases and their order parameters}\label{OPsAll}
%
%
%

%
%
\subsection{Brane operators and correlators}\label{BOPs}

The string operators and string order parameters were generalized for two-leg spin ladders in
\cite{Nishiyama:1995,Watanabe:1995,Shelton:1996,White:1996,Kim:2000,*Fath:2001,*Kim:2008,Nakamura:2003b}.
However it is more consistent to use the concept of the brane order to go beyond one spatial dimension  \cite{NussinovChen:2008,Cirac:2008,Rath:2013,Bahri:2014,Montorsi:2016,*Montorsi:2017,*Montorsi:2019}.

\begin{figure}[h]
\includegraphics[width=8.5 cm]{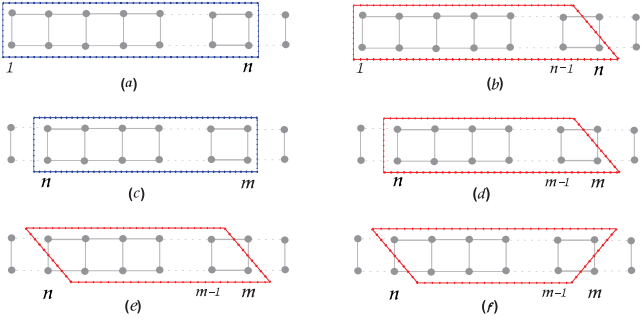}
\caption{Branes and brane-brane correlation functions: (a) and (b) depict an even/odd brane, resp., for (b)
another odd brane may be obtained by reflection with respect to horizontal axis;
(c) -- even-even brane correlator; (d) -- even-odd brane correlator; (e), (f) -- odd-odd brane correlators.
The odd-even and odd-odd brane correlators can also correspond to other graphs obtained form the cases (d-f) by reflections with
respect to horizontal/vertical axes.}
\label{br_op}
\end{figure}

We define an even brane operator which includes the area with an integer number of legs:
\begin{equation}
\mathcal{B}_e^i  (n) \equiv \prod_{\alpha=1}^2 \prod_{l=1}^n \sigma_{\alpha}^i (l),~~~i=x,y,z~,
\label{BOe}
\end{equation}
and the odd brane operator which includes one ``loose" extra spin at the far right end:
\begin{equation}
\mathcal{B}_{o,\alpha}^i  (n) \equiv \mathcal{B}_e^i  (n-1) \sigma_{\alpha}^i (n),~~ \alpha =1,2.
\label{BOo}
\end{equation}
We also define the corresponding brane-brane correlation functions which are calculated in the following: the
even-even correlator
\begin{equation}
\label{Bee}
  \langle \mathcal{B}_e^i  (m) \mathcal{B}_e^i  (n) \rangle~,
\end{equation}
the mixed correlators
\begin{equation}
\label{Beo}
  \langle \mathcal{B}_e^i  (m) \mathcal{B}_{o,\alpha}^i  (n) \rangle~~\mathrm{and}~ e \leftrightarrow o~,
\end{equation}
and the odd-odd correlation function
\begin{equation}
\label{Boo}
  \langle \mathcal{B}_{o,\alpha}^i (m) \mathcal{B}_{o,\beta}^i (n)\rangle~.
\end{equation}
These brane-brane correlators are schematically depicted  in  Fig.~\ref{br_op}. In the following we will also encounter the brane correlation function of
the operator
\begin{equation}
\label{BzXY}
  \mathcal{B}_e^z  (n-1) \sigma_1^x(n) \sigma_2^y(n) ~~\mathrm{and}~ x \leftrightarrow y~,
\end{equation}
which is the even $z$-brane with the $x$ and $y$ spins attached to its far right edge.

The brane order parameters are defined as non-vanishing limits of corresponding brane-brane correlation functions as $m-n \to \infty$.

%
%
%
%

%
%
\subsection{Staggered ladder}\label{StagOPs}

In the previous section we worked out the mean-field approximation for the staggered ladder to map its Hamiltonian onto a sum of
two decoupled dual XY chains \eqref{HoXY}  and \eqref{HeXY}. The phase boundaries \eqref{PhBs} shown in Fig.~\ref{PhaseDiag} are deduced from the
fermionic spectrum  \eqref{spectra_s}. Fig.~\ref{PhaseDiag} can now be more easily reproduced from superimposing on the
$(\delta,\gamma)$ plane the phase diagrams of the even and odd XY chains (\ref{HoXY},\ref{HeXY}). The information on phases and order
parameters of such chain is given in Appendix~\ref{AppB}. The dual order parameters (expressed via $\tau$ and $\mu$ operators)
for each of the nine phases (A-I) shown in Fig.~\ref{PhaseDiag}, are presented in Table~\ref{Table1}.
From  Hamiltonians \eqref{HoXY}  and \eqref{HeXY} we easily establish the symmetry of the staggered ladder:
\begin{eqnarray}
\label{DmD}
  \delta &\mapsto&  -\delta: ~\mathrm{even} \leftrightarrow \mathrm{odd},~ \tau \leftrightarrow \mu \\
\label{GmG}
  \gamma  &\mapsto&  -\gamma: ~\mathrm{even} \leftrightarrow \mathrm{odd},~ \tau \leftrightarrow \mu,~x \leftrightarrow y~.
\end{eqnarray}

\begin{table*}[t]
\begin{center}
\caption{Order parameters in terms of dual and original spins, winding numbers, number of the Majorana zero-energy edge modes $N_{\s M}$
for nine phases of the staggered ladder shown in Fig.~\ref{PhaseDiag}.}
\label{Table1}
\begin{tabular}{|c|c|c|c|c|c|c|c|}
  \hline
  ~Reqion~ & ~$\tau$ order~ & $~\mu$ order~ & $~\sigma$ order~  & $~N_w^e~$  & $~N_w^o~$  & $~N_w~$  & $~N_{\s M}~$\\
  ~~ & ~(odd sector)~ & ~(even sector)~ & ~~ & ~~ & ~~ & ~~ & ~~ \\
  \hline
  A & $\langle\tau_{x}\rangle$ & $\langle\mu_{y}\rangle$ & $\mathcal{O}_{z,3}$ & 1 & -1 & 0 & 4\\
  B & $\langle\tau_{x}\rangle$ & $\mathcal{O}_{z,\mu}$ & $m_x$ & 0 & -1 & -1 & 2 \\
  C & $\langle\tau_{x}\rangle$ & $\langle\mu_{x}\rangle$ & $\mathcal{O}_z |_{x}^{y}$  & -1 & -1 & -2 & 4 \\
  D & $\mathcal{O}_{z,\tau}$ &  $\langle\mu_{x}\rangle$ & $m_x$  & -1 & 0 & -1 & 2 \\
  E & $\langle\tau_{y}\rangle$ & $\langle\mu_{x}\rangle$  & $\mathcal{O}_{z,3}$  & -1 & 1 & 0 & 4 \\
  F & $\langle\tau_{y}\rangle$ & $\mathcal{O}_{z,\mu}$ & $m_y$  & 0 & 1 & 1 & 2 \\
  G & $\langle\tau_{y}\rangle$ & $\langle\mu_{y}\rangle$  & $\mathcal{O}_z |_{y}^{x}$  & 1 & 1 & 2 & 4 \\
  H & $\mathcal{O}_{z,\tau}$ & $\langle\mu_{y}\rangle$ & $m_y$  & 1 & 0 & 1 & 2 \\
  I & $\mathcal{O}_{z,\tau}$ & $\mathcal{O}_{z,\mu}$ & $\mathcal{O}_{z ,1}$  & 0 & 0 & 0 & 0 \\
  \hline
\end{tabular}
\end{center}
\end{table*}

After relabelling ladder's sites according to the path shown in Fig.~\ref{SnakePath}, the brane operators introduced
in the previous subsection can be represented via string operators defined along a given path (say $\mathcal{P}$)
as
\begin{equation}
\label{Oi}
  O_i(n) \equiv  \prod_{\s l \leq n, l \in \mathcal{P}} \sigma^i_{ l},~~~i=x,y,z.
\end{equation}
We also introduce a short-hand notation for the product of two sting operators
\begin{equation}
\label{Dii}
  D_{ii}(L,R) \equiv  O_i(L-1) O_i(R) =\prod_{l=L}^R \sigma^i_l,
\end{equation}
such that its average yields the string-string correlation function
\begin{equation}
\label{Dcorr}
  \mathfrak{D}_{ii}(L,R) \equiv \langle D_{ii}(L,R)\rangle ~.
\end{equation}
Now we will find explicitly the order parameters in four regions (A,B,C,I) of the phase diagram in Fig.~\ref{PhaseDiag},
while the rest of the phases can be done using the symmetry \eqref{DmD} and \eqref{GmG}, without calculations.\\

\begin{figure}[h]
\includegraphics[width=9.0 cm]{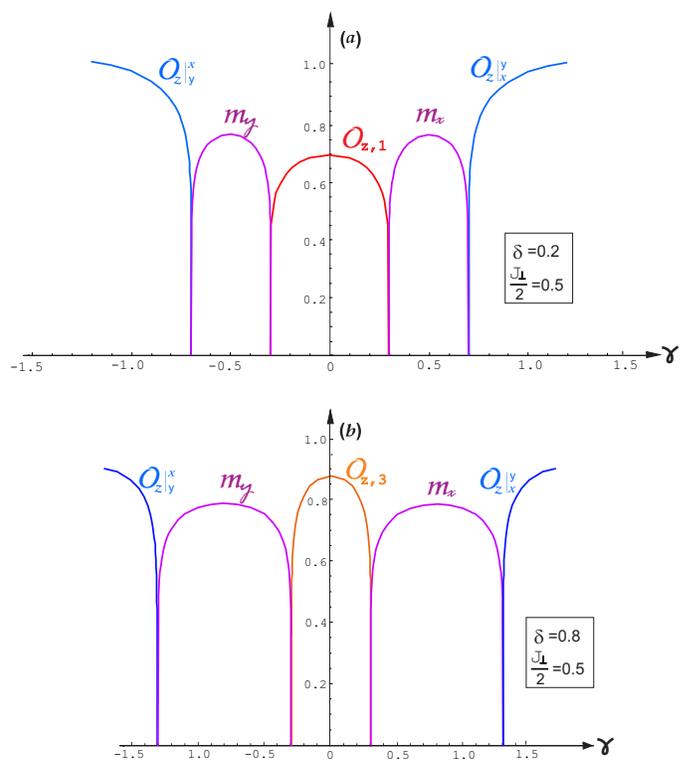}
\caption{Order parameters of the staggered ladder along two paths  indicated in Fig.~\ref{PhaseDiag}.
(a): $\delta =0.2$ (path 1); (b): $\delta=0.8$ (path 2); $t=1$ in both cases.}
\label{OPsStag}
\end{figure}

\textbf{\emph{Region (A):}}
The dual magnetizations are found using \eqref{mxAlt}:
\begin{eqnarray}
\label{taux}
  \langle \tau_{\s 1}^{x}\tau_{\s 2N+1}^{x}\rangle &\xmapsto[\s N \to \infty]& \langle\tau_{x}\rangle^2
  =\frac{2\sqrt{t}}{t+\gamma+\delta}\Big[(\gamma+\delta)^{2}-\frac{J_\perp^{2}}{4}\Big]^{1/4} ~~~~~ \\
 \label{muy}
  \langle \mu_{\s 0}^{y}\mu_{\s 2N}^{y}\rangle &\xmapsto[\s N \to \infty]& \langle\mu_{y}\rangle^2
  =\frac{2\sqrt{t}}{t+|\gamma-\delta|}\Big[(\gamma-\delta)^{2}-\frac{J_\perp^{2}}{4}\Big]^{1/4}~~~~~~
\end{eqnarray}
Using the duality transformations (\ref{XX-ti},\ref{YY-ti},\ref{taumuDual1},\ref{taumuDual2}) we can relate the two-point correlators of the dual spins to the string-string
correlations functions of the spins $\sigma$ as:
\begin{eqnarray}
\label{tauD}
  \tau_{\s 1}^{x}\tau_{\s 2N+1}^{x} &=& D_{xx}(2,2N+1), \\
\label{muD}
   \mu_{\s 0}^{y}\mu_{\s 2N}^{y} &=&  D_{yy}(2,2N+1),
\end{eqnarray}
resulting in the $z$-string order parameter in this region of the phase diagram
\begin{eqnarray}
\label{taumuOz3}
  \langle  \tau_{\s 1}^x \tau_{\s 2N+1}^x \mu_{\s 0}^y \mu_{\s 2N}^y \rangle &=&
  (-1)^N \mathfrak{D}_{zz}(2,2N+1) \xmapsto[\s N \to \infty]  ~\pm \mathcal{O}_{z,3}^2 \nonumber \\
  \mathcal{O}_{z,3} &=& \langle\tau_{x}\rangle \langle\mu_{y}\rangle~.
\end{eqnarray}
Note that the above string correlator is equivalent to the odd-odd $z$-brane correlation function \eqref{Boo}
shown in Fig.~\ref{br_op}:
$\mathfrak{D}_{zz} \leftrightarrows \langle \mathcal{B}_{o,\alpha}^z  \mathcal{B}_{o,\beta}^z \rangle$,
so $\mathcal{O}_{z,3}$ defined by Eqs.~(\ref{taumuOz3},\ref{taux},\ref{muy}) is an exact value of the brane order
parameter. It is plotted in Fig.~\ref{OPsStag}(b) along the path (2) indicated in Fig.~\ref{PhaseDiag}.\\

\textbf{\emph{Region (B):}} The order parameter of $\mu^z$ strings residing in the even
sector is readily obtained from Eq.~\eqref{OzAnSt}:
\begin{eqnarray}
\label{Ozmu1}
 \Big\langle \prod_{l=0}^N \mu^z_{\s 2l} \Big\rangle  \xmapsto[\s N \to \infty]  ~\mathcal{O}_{z,\mu}^2
  =  \Big[ \frac{J_\perp^2/4-(\gamma-\delta)^2}{J_\perp^2/4-(\gamma-\delta)^2+t^2} \Big]^{1/4}~~~
\end{eqnarray}
The dualities (\ref{XX-ti},\ref{YY-ti},\ref{taumuDual1},\ref{taumuDual2}) yield
\begin{equation}
\label{OzmuDxx}
  \prod_{l=0}^{N} \mu^z_{\s 2l}=D_{xx}(1,2N+2))~,
\end{equation}
so the overlapping dual orders amount to a conventional antiferromagnetic order of the original spins $\sigma$ in this phase:
\begin{equation}
\label{Dualmx}
  \langle \tau_{\s 1}^x \tau_{\s 2N+1}^x  \prod_{l=0}^{N} \mu^z_{\s 2l} \rangle= \langle  \sigma_{\s 1}^x \sigma_{\s 2N+2}^x \rangle
  \xmapsto[\s N \to \infty] ~\pm m_x^2~,
\end{equation}
where the explicit formula for magnetization $m_x=\langle\tau_{x}\rangle \mathcal{O}_{z,\mu}$ is given by Eqs.~(\ref{taux},\ref{Ozmu1}).\\

\textbf{\emph{Region (C):}} The dual magnetization is readily found in this case as:
\begin{eqnarray}
 \label{mux}
  \langle \mu_{\s 0}^{x}\mu_{\s 2N}^{x}\rangle &\xmapsto[\s N \to \infty]& \langle\mu_{x}\rangle^2
  =\frac{2\sqrt{t}}{t+\gamma-\delta}\big[(\gamma-\delta)^{2}-\frac{J_\perp^{2}}{4}\big]^{1/4}~~~~~~
\end{eqnarray}
We use the duality transformations to find
\begin{equation}
\label{muxsigma}
  \mu_{\s 0}^{x}\mu_{\s 2N}^{x}= \sigma_{\s 1}^x \sigma_{\s 2}^x D_{yy}(2,2N+1) \sigma_{\s 2N+1}^x \sigma_{\s 2N+2}^x~,
\end{equation}
and then we define the order parameter of this phase from the limit of the overlapping dual correlators as
\begin{eqnarray}
\label{taumuOzxy}
  \langle  \tau_{\s 1}^x \tau_{\s 2N+1}^x \mu_{\s 0}^x \mu_{\s 2N}^x \rangle &=&
  (-1)^N \langle  \sigma_{\s 1}^x \sigma_{\s 2}^y D_{zz}(3,2N)\sigma_{\s 2N+1}^y \sigma_{\s 2N+2}^x \rangle  \nonumber \\
   &\xmapsto[\s N \to \infty]& ~\pm \big( \mathcal{O}_z |_{x}^{y} \big)^2~, \nonumber \\
   \mathcal{O}_z |_{x}^{y} &=& \langle\tau_{x}\rangle \langle\mu_{x}\rangle~.
\end{eqnarray}
Such quite tricky order parameter $\mathcal{O}_z |_{x}^{y}$ is obtained from the limit of the correlation function of two $z$-stings
which both have $\sigma^y$ and $\sigma^x$ spins attached to their right ends. Those strings with attachments are the chain representations of the
brane operators with edge attachments defined by \eqref{BzXY}. Curiously enough, such tricky order parameter is given by a simple analytical expression
from  Eqs.~(\ref{taux},\ref{mux}).
\cite{[{Similar ``edged'' brane order parameters were reported in the Kitaev-Heisenberg two-leg ladder (an example of anisotropic ladder resembling the present model)  in the numerical work by~}] Sorensen:2019,*Agrapidis:2019}
\\

\textbf{\emph{Region (I):}} We use again Eq.~\eqref{OzAnSt} from Appendix to calculate the $\tau ^z$-string order parameter:
\begin{eqnarray}
\label{Oztau1}
 \Big\langle \prod_{l=1}^{N+1} \tau^z_{\s 2l-1} \Big\rangle  \xmapsto[\s N \to \infty]  ~\mathcal{O}_{z,\tau}^2
  =  \Big[ \frac{J_\perp^2/4-(\gamma+\delta)^2}{J_\perp^2/4-(\gamma+\delta)^2+t^2} \Big]^{1/4}~~~
\end{eqnarray}
and the duality transformations to get
\begin{equation}
\label{OzmuDyy}
  \prod_{l=1}^{N+1} \tau^z_{\s 2l-1}=D_{yy}(1,2N+2)~.
\end{equation}
The order parameter for this phase is found then as
\begin{eqnarray}
\label{OtaumuOz1}
  \Big\langle   \prod_{l=1}^{N+1} \tau^z_{\s 2l-1} \mu^z_{\s 2l-2} \Big\rangle
  &=& (-1)^{N+1} \mathfrak{D}_{zz}(1,2N+2) \xmapsto[\s N \to \infty] ~\pm\mathcal{O}_{z,1}^2 \nonumber \\
  \mathcal{O}_{z,1} &=&\mathcal{O}_{z,\tau}\mathcal{O}_{z,\mu} ~,
\end{eqnarray}
with the explicit expression given by \eqref{Ozmu1} and \eqref{Oztau1}.

Thus, all order parameters for each phase of the phase diagram of the staggered ladder are found analytically
as closed expressions in terms of the \textit{renormalized} couplings of the model.  To visualize the results of this subsection,
the local and non-local (brane) order parameters are plotted in Fig.~\ref{OPsStag} (a,b) along two paths indicated in Fig.~\ref{PhaseDiag}.
Note that in the limit $J_\perp \to 0$ the antiferromagmetic phases shown in Fig.~\ref{PhaseDiag} vanish, although single chains are known
to be antiferromagnetically orderd in the regions (C) and (G) \cite{Chitov:2017JSM}. This proves that the brane order parameters probe the long-ranged order in both directions: along the chains and along the rungs.

%
%
%
%

\subsection{Columnar ladder}\label{ColOPs}

The effective Hamiltonian of the columnar ladder is mapped onto two equivalent dimerized XY chains in a staggered transverse field (\ref{HoXYCol},\ref{HeXYCol}). Contrary to the previous case, the columnar ladder has identical spectra in the even and odd sectors, resulting in the two-fold degeneracy of the eigenvalues of the Hamiltonisan \eqref{spectra_c}. As a consequence, the phase diagram of this case is simpler, since there are no regions in the parametric space where different orders in the even and odd sectors overlap, resulting in larger variety of phases in Fig.~\ref{PhaseDiag}. The phase diagram of the columnar ladder consists of only three phases of the chain \eqref{HoXYCol} (or \eqref{HeXYCol}), where the same even/odd order parameters coexist, see
Fig.~\ref{Phasecol}.

The region (I) of the phase diagram in Fig.~\ref{Phasecol} has its the counterpart on the phase diagram of the staggered ladder,
discussed above in detail. The string order parameter  of this phase is:
\begin{equation}
\label{Oz1Col}
  \mathcal{O}_{z,1} =\mathcal{O}_{z,\tau}\mathcal{O}_{z,\mu}=\mathcal{O}_{z,\tau}^2
\end{equation}
The order parameter in the phase (C) is the brane parameter with the edge spin attachments:
\begin{equation}
\label{OzAttCol}
  \mathcal{O}_z |_{x}^{y} =\langle\tau_{x}\rangle \langle\mu_{x}\rangle=\langle\tau_{x}\rangle^2~,
\end{equation}
and for the phase (G) the parameters are obtained from $x \leftrightarrows y$. The equivalent even and odd sectors of the dual Hamiltonian of the columnar ladder lead not only to the degeneracy of the spectrum, but also, as one can see from (\ref{Oz1Col},\ref{OzAttCol}) to the change of the universality class on the lines of continuous phase transitions  (bold red lines in
Fig.~\ref{Phasecol}). In the effective free-fermionic approximation used in this paper, the order parameters of all phases
of the staggered ladder have the critical index $\beta =1/8$ (2D Ising universality class), while the critical index of the order
parameters (\ref{Oz1Col},\ref{OzAttCol}) of the columnar ladder is twice the 2D Ising value. From the scaling relations we find the critical indices
(universality class) of the columnar ladder in the free-fermionic approximation:
\begin{eqnarray}
  \beta &=& 1/4,~~ \nu=1,~~ \alpha =0, \nonumber \\
  \eta &=& 1/2,~~ \gamma =3/2.
\label{UnivCol}
\end{eqnarray}

\begin{figure}[h]
\includegraphics[width=7.0 cm]{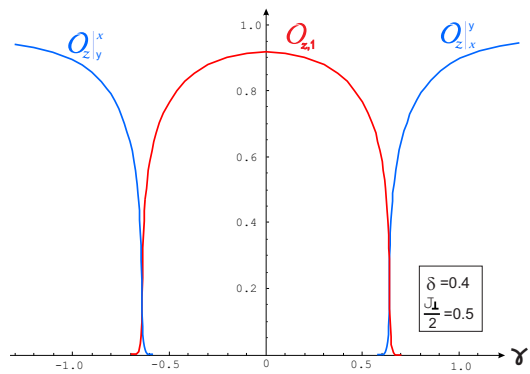}
\caption{Brane order parameters in the columnar ladder along the path indicated in Fig.~\ref{Phasecol}.
The tails seen near the critical points are finite-size effects:
the results are obtained numerically from $140 \times 140$ Toeplitz matrices.}
\label{OPsCol}
\end{figure}

For the dual chains (\ref{HoXYCol},\ref{HeXYCol}) with dimerization no analytical results for the magnetization or the oscillating string order parameter are available. The parameters  $\langle\tau_{x}\rangle$ and $\mathcal{O}_{z,\tau}$ are calculated numerically from the Toeplitz determinants, using the results for the generating functions and the Toeplitz matrices given in  \cite{Chitov:2019}. The brane order parameters calculated from the Toeplitz matrices of the size $140 \times 140$ are plotted in Fig.~\ref{OPsCol}. The calculations are done along the path indicated in Fig.~\ref{Phasecol}.

%
%
%
%

%
%
\section{Beyond Landau: winding numbers, Majoranas, and disentanglement}\label{TopNum}
%
%
%
%
In this section we analyze and calculate several quantities which are not a part
of the standard Landau framework. However, they provide complementary information
helpful to sharpen our understanding of criticality.

In the mean-field approximation of the effective fermionic Hamiltonian (free-fermionic approximation),
all quantum phase transitions of the staggered ladder shown in Fig.~\ref{PhaseDiag} belong to the 2D
Ising universality class. The gap equation \eqref{GapSt} leads to the critical index $\nu=1$; the local
and brane order parameters calculated in Sec.~\ref{StagOPs} have their critical index $\beta =1/8$.
This implies the critical index of the specific heat $\alpha =0$. The universality class of the quantum
phase transitions in the columnar ladder shown in Fig.~\ref{Phasecol}, is different, cf. Eqs.~\eqref{UnivCol}.
However the index $\alpha=0$ for both types of the dimerized ladders.

The singular behavior of specific heat, i.e., the derivative of the entropy near the thermal phase transition
of the second kind, has its counterpart for the continuous quantum phase transition at $T=0$, and is related to
the entanglement. For reviews, see, e.g.,  \cite{Amico:2008,*Laflorencie:2016}. In the context of the present study
it is most convenient to use the global entanglement $\mathcal{E}$  \cite{WeiGold:2005,*WeiGold:2011}.
It measures proximity of the given quantum state to a probe factorized (disentangled or ``classical") state.
As follows from the results of Wei \textit{et al} for the free-fermionic Hamiltonian \cite{WeiGold:2005,*WeiGold:2011},
the derivative of the global entanglement diverges with the critical index of the specific heat ($\alpha=0$)
\begin{equation}
\label{dEh}
  \mathcal{E}^\prime (\epsilon) \propto - \ln |\epsilon|,~ |\epsilon| \ll 1~
\end{equation}
while approaching quantum phase transitions shown in Figs.~\ref{PhaseDiag} and \ref{Phasecol}. Here $\epsilon$
stands for the parametric distance to the quantum critical point, cf. Eqs.~\eqref{GapSt} and \eqref{DelCol}.

The other important property of the global entanglement is that it vanishes ($\mathcal{E}=0$) in the factorized ground state.
The latter is also characterized by vanishing pairwise concurrence $\mathcal{C}=0$ \cite{Amico:2006},
which is another measure of entanglement introduced by Wootters \cite{Wootters:1998}.

\subsection{Staggered ladder}\label{StagDES}
%
%
%
\subsubsection{Winding numbers and Majorana modes}
%
%
%

After mapping of the effective fermionic Hamiltonian of the staggered ladder onto the pair
of decoupled XY chains (\ref{HoXY},\ref{HeXY}), the analysis  can be done using the results for the $XY$ chain in
transverse fields  \cite{Chitov:2019,Chitov:2022}. The four eigenvalues  $E_{\pm\pm}(k)$ \eqref{spectra_s} found
from diagonalization of the $8\times 8$ Hamiltonian matrix \eqref{HAB},
correspond to the two pairs of eigenvalues of the anisotropic even and odd $XY$ chains with the anisotropy parameters
\begin{equation}
\label{gameo}
  \gamma_{o/e}=\gamma \pm \delta
\end{equation}
and the alternating transverse field
\begin{equation}
\label{haeo}
  h_a^e=h_a^o=J_\perp/2~.
\end{equation}
The effective dimerization and the uniform transverse field are absent
\begin{equation}
  \delta_e = \delta_o =0, ~~
  h^e = h^o=0
\label{hdStag}
\end{equation}
in the even and odd dual Hamiltonians  (\ref{HoXY},\ref{HeXY}).

The fermionic representation of the even/odd sector of the staggered Hamiltonian yields a $4\times 4$ matrix analogous to \eqref{HAB} with
$2 \times 2$ blocks \cite{Chitov:2019,Chitov:2020}:
\begin{equation}
\label{A2}
  \hat{A} \equiv  \left(%
\begin{array}{cc}
  t \cos k & J_\perp/2 \\
 J_\perp/2 &  -t \cos k \\
\end{array}%
\right)~,
\end{equation}
and
\begin{equation}
\label{B2}
 \hat{B}_{\s \sharp} \equiv  \left(%
\begin{array}{cc}
  -i \gamma_{\s \sharp} \sin k & 0 \\
  0 & i \gamma_{\s \sharp} \sin k \\
\end{array}%
\right)~,~~\sharp=e,o.
\end{equation}
For the Hamiltonian of this type the winding number is defined as the following integral over the Brillouin zone \cite{Schnyder:2008,SchnyderRyu:2011}
\begin{equation}
\label{Nw}
  N_w^{\s \sharp}=\frac{1}{2\pi i} \int_{-\pi/2}^{\pi/2} dk \partial_k \ln \mathrm{det}\hat{D}_{\s \sharp}~,
\end{equation}
where $\hat{D}_{\s \sharp}(k) \equiv \hat{A} (k)+\hat{B}_{\s \sharp}(k)$. To avoid ambiguities related to the definition of the phase of
$\mathrm{det}\hat{D}_{\s \sharp}(k)$ at the ends of the Brillouin zone \cite{Chitov:2019}, one needs to check whether the path of $\mathrm{det}\hat{D}_{\s \sharp}(k)$
on the complex plane encloses the origin during the integration.  The latter happens if  $\mathrm{det}\hat{D}_{\s \sharp}(\pm \pi/2)$ and
$\mathrm{det}\hat{D}_{\s \sharp}(0)$ have opposite signs. An explicit calculation with \eqref{A2} and \eqref{B2} yields
\cite{[{This formula is a special case of the result for $N_w$ derived by A. Furusaki. We thank him for an exchange about this issue and for
sharing with us his unpublished notes: }] Furusaki}:
\begin{equation}
\label{NweoEq}
  N_w^{\s \sharp}=-\mathrm{sign}(\gamma_{\s \sharp})\Theta(\gamma_{\s \sharp}^2-J_\perp^2/4)~,
\end{equation}
where $\Theta(x)$ is the Heaviside step function. The winding number for the ladder is found as
\begin{equation}
\label{Nwsum}
  N_w= N_w^e +N_w^o~,
\end{equation}
and its values for each phase are given in Table~\ref{Table1}.

Another angle of analysis can be presented upon analytical continuation of the wave numbers onto the complex plane:
$e^{i 2 k} =z$. Then the winding number \eqref{Nw} becomes the logarithmic residue of $\mathrm{det}\hat{D}_{\s \sharp}$ :
\begin{equation}
\label{NwCompl}
    N_w^{\s \sharp}=\oint _{\left|z\right|=1} \frac{dz}{2\pi i} \partial_z \ln  \mathrm{det}\hat{D}_{\s \sharp}~.
\end{equation}
It accounts for the excess of the number of zeros over the number of poles (weighted with their degrees of multiplicity)
of $\mathrm{det}\hat{D}_{\s \sharp}$ inside the unit circle on the complex plane \cite{Whittaker:1915}. The zeros of $\mathrm{det}\hat{D}$
are also zeros of the spectra $E_\pm$ of the XY chain, since $(E_+ E_-)^2=\mathrm{det}\hat{D}\hat{D}^\dag$.
Any change of winding number means that a root (roots) crossed the unit circle $|z|=1$, which signals a quantum phase transition
\cite{Verresen:2018,Chitov:2021}. For the Hamiltonians (\ref{HoXY},\ref{HeXY}) we find
\begin{equation}
\label{detD}
  \mathrm{det}\hat{D}_{\s \sharp}(z)=-\frac{(1-\gamma_{\s \sharp})^2}{4z}(z-z_+)(z-z_-)~,
\end{equation}
where $z_\pm$ are the roots of the quadratic equation
\begin{equation}
\label{zeqn}
  (1-\gamma_{\s \sharp})^2 z^2 +2(1-\gamma_{\s \sharp}^2+J_\perp^2/2) z+(1+\gamma_{\s \sharp})^2=0~,
\end{equation}
whence the results of Table~\ref{Table1} are recovered.

The important property of the roots $z_\pm$ is that they are also the eigenvalues of the transfer matrix which yields the wave functions of the zero-energy
Majorana modes localized near the opposite ends of the dual (even or odd) chains \cite{Chitov:2022}. More details and the explicit formulas for those
wave functions can be found in \cite{Chitov:2022}. The qualitative result is that for each phase in the even or odd sector with $N_w^{\sharp}=\pm 1$, there are two edge modes in the corresponding sector. With the available winding numbers $N_w^{\sharp}$, one easily obtains the total number of the Majorana edge modes $N_{\s M}$ for each phase, see Table~\ref{Table1}.

%
%
\subsubsection{Disorder lines and disentanglement}
%
%
%

It was shown recently \cite{Chitov:2022} that the dual chains (\ref{HoXY},\ref{HeXY},\ref{HoXYCol},\ref{HeXYCol}) manifest also  a special type of weak transition, called disorder lines (DLs) by Stephenson \cite{Stephenson-I:1970,*Stephenson-II:1970,*Stephenson:1970PRB}. The transition consists in modulation of the exponential decay of correlation functions and the wave functions of the Majorana zero modes by the incommensurate (IC) oscillations.
Using the correspondence between the parameters of the original ladder and the dual chains  (\ref{gameo},\ref{haeo},\ref{hdStag}) in the equations for the
DLs given in \cite{Chitov:2022}, we readily find four straight DLs in the staggered ladder:
\begin{equation}
\label{DLst}
  \gamma_{\s \sharp}^2=t^2+J_\perp^2/4,
\end{equation}
shown in the phase diagram,  see Fig.~\ref{PhaseDiag}.

The DLs \eqref{DLst} bound the regions on the phase diagram where the functions of the \textit{even sector} manifest the IC oscillations at:
\begin{equation}
\label{ICe}
 |\gamma| >  \Big| \delta \pm \sqrt{t^2+J_\perp^2/4} \Big|.
\end{equation}
The IC oscillating regions for their counterparts of the \textit{odd sector} are localized at:
\begin{equation}
\label{ICo}
 |\gamma| >  \Big| -\delta \pm \sqrt{t^2+J_\perp^2/4} \Big|.
\end{equation}
The wave numbers  $q_{\s \sharp}$ of the IC oscillations are defined in the reciprocal space of the dual even/odd chains. From the results of \cite{Chitov:2022} two distinct even/odd IC wave numbers are found:
\begin{equation}
\label{qeo}
  q_{\s \sharp}=  \arcsin\frac{J_\perp}{2\sqrt{\gamma_{\s \sharp}^2-t^2}}~,
\end{equation}
where $q_{\s \sharp}$ evolves smoothly from $\pi/2$ on the DL to $q_{\s \sharp} \to 0$ when $\gamma_{\s \sharp} \to \infty$.

Since the effective Hamiltonian is a sum of two commuting even and odd terms (\ref{HoXY},\ref{HeXY}), its ground state is a direct product
of the even and odd vacuum states:
\begin{equation}
\label{GSpr}
  |GS \rangle = |GS_e \rangle \otimes  |GS_o \rangle~.
\end{equation}
DLs of the even/odd Hamitonians (\ref{HoXY},\ref{HeXY}) are also the special disentanglement points where their corresponding ground states
$|GS_{e/o} \rangle$ are factorized \cite{Chitov:2022}. Instead of using the results of \cite{Chitov:2022} for the special case  (\ref{gameo},\ref{haeo},\ref{hdStag}),
it is technically simpler to apply the duality transformation discussed in Appendix B, and map (\ref{HoXY},\ref{HeXY}) onto the XY chains with uniform
field. The latter model has the well-known factorized ground state, see, e.g.,  \cite{Franchini:2017}. Then we readily write the factorized even/odd
states on the DLs \eqref{DLst}:
\begin{equation}
\label{GSeo}
  |GS_{\sharp} \rangle =
  \prod _{n=1}^{N}\left(\cos \vartheta_{\sharp}~ {\left| \uparrow \right\rangle_{2n/2n-1}  } + \sin \vartheta_{\sharp}~ {\left| \downarrow \right\rangle_{2n/2n-1}  } \right)~,
\end{equation}
where the even/odd parametric angles $\vartheta_{\sharp}$ are defined via the following equation:
\begin{equation}
\label{Coseo}
  \cos^2 2\vartheta_{\sharp}= \frac{|\gamma_{\s \sharp}|-t}{|\gamma_{\s \sharp}|+t}~.
\end{equation}
As a consequence of the factorization of the even or odd component of the GS \eqref{GSeo}, the even/odd two-point correlation functions
defined on the dual sites are strictly constant on their corresponding DLs \eqref{DLst} (\textit{up to an obvious antiferromagnetic sign alternation,
implicitly presumed throughout}).

Below we provide several examples of the correlators in two regions (A) and (C) of the phase diagram in Fig.~\ref{PhaseDiag}, while the other
regions can be worked out using the symmetry \eqref{DmD} and \eqref{GmG}.

\textbf{\emph{Region (A):}} Using $|GS_o \rangle$ given by Eq.~\eqref{GSeo} on the \textit{odd} DL in the calculations of average quantities, we find:
\begin{eqnarray}
\label{txO}
  \langle \tau_{\s 2n+1}^{x} \rangle  &=&  \pm \sin 2\vartheta_o,~
  \langle \tau_{\s 2m+1}^{x}\tau_{\s 2n+1}^{x}\rangle = \pm \sin^2 2\vartheta_o,\\
\label{tzO}
  \langle \tau_{\s 2n+1}^{z} \rangle &=& \pm \cos 2\vartheta_o,~
  \langle \tau_{\s 2m+1}^{z}\tau_{\s 2n+1}^{z}\rangle =\pm \cos^2 2\vartheta_o, \\
\label{tyO}
 \langle \tau_{\s 2n+1}^{y} \rangle  &=&   \langle \tau_{\s 2m+1}^{y}\tau_{\s 2n+1}^{y}\rangle =0,~~~ \forall~ n \neq m~.
\end{eqnarray}
The last equation can be also understood as a consequence of the ``classicality'' of the GS vector $|GS_o \rangle$:
\begin{equation}
\label{SumO}
  \langle \tau_{\s 2n+1}^{x} \rangle^2+\langle \tau_{\s 2n+1}^{z} \rangle^2=
\sin^2 2\vartheta_o+\cos^2 2\vartheta_o=1,~ \forall~ n~.
\end{equation}
Using above results and relation \eqref{tauD} for the string of the original spins $\sigma^x$ in the ladder, we obtain
the constant string correlation function
\begin{equation}
\label{DxxO}
  \mathfrak{D}_{xx}(2,2n+1)= \pm \sin^2 2\vartheta_o,~ \forall~ n~.
\end{equation}
Using $|GS_e \rangle$ given by Eq.~\eqref{GSeo}, we find the constant dual correlation functions on the \textit{even} DL:
\begin{eqnarray}
\label{muyE}
  \langle \mu_{\s 2n}^{y} \rangle  &=& \pm \sin 2\vartheta_e,~
  \langle \mu_{\s 2m}^{y}\mu_{\s 2n}^{y}\rangle =\pm \sin^2 2\vartheta_e,\\
\label{muzE}
  \langle \mu_{\s 2n}^{z} \rangle &=& \pm \cos 2\vartheta_e,~
  \langle \mu_{\s 2m}^{z}\mu_{\s 2n}^{z}\rangle =\pm \cos^2 2\vartheta_e, \\
\label{muxE}
 \langle \mu_{\s 2n}^{x} \rangle  &=&   \langle \mu_{\s 2m}^{x}\mu_{\s 2n}^{x}\rangle =0,~~~ \forall~ n \neq m~,
\end{eqnarray}
with
\begin{equation}
\label{SumE}
  \langle \mu_{\s 2n}^{y} \rangle^2+\langle \mu_{\s 2n}^{z} \rangle^2=
\sin^2 2\vartheta_e+\cos^2 2\vartheta_e=1,~ \forall~ n~,
\end{equation}
and constant correlation function of the strings of ladder's  spins $\sigma^y$, cf. \eqref{muD}:
\begin{equation}
\label{DyyE}
  \mathfrak{D}_{yy}(2,2n+1)= \pm \sin^2 2\vartheta_e,~ \forall~ n~.
\end{equation}
The constant correlation functions are a hallmark of disentanglement. However, the above results imply only the partial disentanglement on the DLs due to
factorization of the even or odd sectors of the GS  \eqref{GSpr}. Consequently, concurrence $\mathcal{C}$ or global entanglement $\mathcal{E}$ do not vanish on the DLs \eqref{DLst},
since the contributions to those quantities from the even and odd sectors mix up. Instead these quantities attain their minima:
\begin{equation}
\label{ECmin}
  \mathrm{even/odd~DLs:}~~\mathcal{C},\mathcal{E} \mapsto \mathrm{min}.
\end{equation}
The complete factorization of the GS \eqref{GSpr} occurs when the even and odd DLs \eqref{DLst} cross:
\begin{equation}
\label{DLcrossA}
\mathrm{Disentanglement:~}\gamma = 0,~ \delta^2=t^2+J_\perp^2/4~.
\end{equation}
At this point the correlation functions involving both the even and odd dual sites are constant. For instance, the $zz$-string correlation
function \eqref{taumuOz3}
\begin{eqnarray}
\label{DESeo}
 \mathfrak{D}_{zz}(2,2n+1)
  = (-1)^n \langle  \tau_{\s 1}^x \tau_{\s 2n+1}^x \mu_{\s 0}^y \mu_{\s 2n}^y \rangle = \nonumber \\
  = \pm \sin^2 2\vartheta_e \sin^2 2\vartheta_o= \pm  \Big(\frac{2t}{\delta+t} \Big)^2~
\end{eqnarray}
is constant and equal to the order parameter $\pm \mathcal{O}_{z,3}^2$ evaluated at the special point \eqref{DLcrossA} where the entanglement vanishes
\cite{Amico:2006,Chitov:2021,Chitov:2022,WeiGold:2005,*WeiGold:2011}:
\begin{equation}
\label{En0}
 \mathcal{C} = \mathcal{E}=0.
\end{equation}
A useful cross-check: the average quantities calculated in this subsection using the factorized states $|GS_{\sharp} \rangle$  \eqref{GSeo}, coincide with their counterparts from Sec.~\ref{StagOPs}, evaluated on the special lines \eqref{DLst} with the formulas obtained via the duality transformations and asymptotic limits of the Toeplitz determinants, as it must be.

\textbf{\emph{Region (C):}} As the odd DL goes through the regions (A) and (C) of the phase diagram in Fig.~\ref{PhaseDiag},
the results  for the odd sector are given by the same Eqs.~(\ref{txO}-\ref{DxxO}). For the average quantities along the even DL
in region (C), we need to interchange $\mu^x \leftrightarrows \mu^y$ in Eqs.~(\ref{muyE}-\ref{SumE}) for the even sector.
Again, the even/odd DLs in this region correspond to the partial disentanglement and the minima \eqref{ECmin} of the entanglement measures.

The complete factorization of the GS and the disentanglement \eqref{En0} occurs at the point where the even and odd DLs \eqref{DLst} cross:
\begin{equation}
\label{DLcrossC}
\mathrm{Disentanglement:~}\delta = 0,~ \gamma^2=t^2+J_\perp^2/4~.
\end{equation}
At this point the correlation functions are constant. For instance, the correlator of the $z$-brane  with the $(x,y)$-spins at its edge \eqref{taumuOzxy}
\begin{eqnarray}
 \label{DzxyDL}
  \langle  \sigma_{\s 1}^x \sigma_{\s 2}^y D_{zz}(3,2n)\sigma_{\s 2n+1}^y \sigma_{\s 2n+2}^x \rangle
  &=& (-1)^n \langle  \tau_{\s 1}^x \tau_{\s 2n+1}^x \mu_{\s 0}^x \mu_{\s 2n}^x \rangle =\nonumber \\
  = \pm \sin^2 2\vartheta_e \sin^2 2\vartheta_o &=&  \pm  \Big(\frac{2t}{\gamma+t} \Big)^2~
\end{eqnarray}
is constant and equal to the order parameter $\pm \mathcal{O}_z |_{x}^{y}$  evaluated at the disentanglement  point \eqref{DLcrossC}.

%
%
%
\subsection{Columnar ladder}\label{ColDES}
%
%
%

The inverse JW transformation of the effective fermionic Hamiltonian of the columnar ladder yields a pair
of the equivalent decoupled XY chains (\ref{HoXYCol},\ref{HeXYCol}) with the same anisotropies, dimerizations,
alternating transverse fields
\begin{equation}
\label{gdhaCol}
  \gamma_{o}=\gamma_{e}=\gamma~,~ \delta_o=\delta_e=\delta~,~ h_a^e=h_a^o=J_\perp/2~,
\end{equation}
and zero uniform transverse fields
\begin{equation}
  h^e = h^o=0~.
\label{hCol}
\end{equation}
The degenerate spectrum $\pm E_{\pm}(k)$ \eqref{spectra_c} of the $8\times 8$ Hamiltonian matrix
of the columnar ladder corresponds to the energy eigenvalues of these two equivalent chains.

The contributions to the winding number \eqref{Nwsum} from the even and odd sectors
(\ref{HoXYCol},\ref{HeXYCol}) trivially doubles. Using either the $2 \times 2$ matrices $\hat{A}$ and $\hat{B}$ are given
in \cite{Chitov:2019} with $h_a=J_\perp/2$ to evaluate \eqref{Nw}, or the complex formalism  \eqref{NwCompl} with
the roots $z_\pm$ provided in \cite{Chitov:2022}, we end up with the winding and Majorana numbers for the three phases
(C,I,G) shown in Fig.~\ref{Phasecol}, equal to their counterparts of the staggered ladder and given in Table~\ref{Table1}.

From the results for the modulated XY chains \cite{Chitov:2022} we readily find two DLs defined by the hyperbolas:
\begin{equation}
\label{DLCol}
  \gamma = \pm \sqrt{1+4t^2/J_\perp^2}\sqrt{\delta^2+J_\perp^2/4}~,
\end{equation}
shown in the phase diagram,  Fig.~\ref{Phasecol}.  These lines signal the simultaneous appearance of oscillations in the even and
odd sectors of the Hamiltonian, deep in the phases (C) and (G).\footnote{\label{DL2} For the ferromagnetic couplings, the so-called DL of the second kind appears in a certain range of parameters \cite{Chitov:2022}, but we will not discuss this case here.}
The IC wave numbers of oscillations are defined in the reciprocal space of the dual chains as
\begin{equation}
 \label{qICcol}
  q= \arctan \sqrt{\frac{1- \xi}{1+ \xi}}~,
\end{equation}
where
\begin{equation}
\label{x}
\xi \equiv \frac{\sqrt{\gamma^2-(t+\delta)^2} \sqrt{\gamma^2-(t-\delta)^2} }{\gamma^2 -(t^2+\delta^2+J_\perp^2/2)}~.
\end{equation}
Similar to the staggered case, the modulation $q$ evolves smoothly from $\pi/2$ on the DL ($\xi=-1$) to $q \to 0$ ($\xi \to 1$) when $\gamma \to \infty$.

As follows from the results \cite{Chitov:2022}, the DLs of the dimerized XY chain are always entangled, thus
\begin{equation}
\label{ECcol}
 \mathrm{At~}\delta \neq 0:~\mathcal{C}\neq 0,~~\mathcal{E} \neq 0.
\end{equation}
The disentanglement occurs only at the point $\delta=0$ on the DLs, which corresponds to the case \eqref{DLcrossC} analysed above.

%
%
%
\section{Conclusion}\label{Concl}
%
%
%

Two-leg spin-$\frac12$ ladders with anisotropy and two different dimerization patterns are analyzed at zero temperature. After fermionization done via the Jordan-Wigner transformation, the spin model becomes equivalent to the ladder of interacting spinless fermions, known also as the Kitaev ladder. The interacting fermionic model is treated within the Hartree-Fock mean-field approximation which allows us to obtain an effective quadratic fermionic Hamiltonian.  Its renormalized parameters are related to
the bare couplings of the microscopic Hamiltonian via the self-consistent mean-field equations which need to be solved numerically. (Some examples of the numerical solutions are given in  Appendix~\ref{AppA}.) The effective mean-field Hamiltonian is further transformed into a sum of two decoupled Majorana Hamiltonians, which in their turn are mapped via an inverse Jordan-Wigner transformation onto a sum of two even/odd XY quantum chains in the alternating transverse fields. The decoupled Hamiltonians $H_{e,o}$ ($H=H_e+H_o$) commute, so the averaging in the even and odd sectors factorizes.

The ground-state phase diagram of the ladder follows straightforwardly from solutions for zeros of the eigenvalues of the effective fermionic Hamiltonian
in the parametric space for each of the two dimerization patterns. The same results are obtained with more physical insight from the mapping of the ladder
onto a couple of dual spin chains. The analysis is based on our understanding of the spectrum and the ground-state phase diagram of the XY chain in the alternating transverse field: the diagram consists of two conventional magnetic phases and a phase with non-local oscillating string order \cite{Chitov:2019}. The phase diagram of the staggered ladder follows then from superposition of the orders in the dual even/odd  XY chains.
In the case of columnar dimerization the two dual XY chains are equivalent, so the columnar ladder has an extra degeneracy of the spectrum, different universality class, and less rich phase diagram which coincides with that of a single dual (even or odd) chain.

To explore the physical nature of the quantum phases predicted to occur in the anisotropic spin ladders, we introduced and calculated the corresponding brane order parameters. The duality between the effective mean-field Hamiltonian of the ladders and the pair of decoupled XY chains allowed us to calculate the brane order parameter as a product of two independent order parameters in the even/odd XY chains.

The ground-state phase diagram of the staggered ladder contains nine phases,
four of which are really distinct, while the other five can be obtained from the symmetries of the model. The order parameters for all phases are collected in
Table~\ref{Table1}. It is straightforward to establish how breaking of the hidden $\mathbb{Z}_2 \otimes \mathbb{Z}_2$ symmetry   \cite{Oshikawa:1992,*Kennedy:1992,*Kohmoto:1992} of the spin ladder in each of the phases is related to the symmetry breaking in the even or odd dual spin chains.
The local order parameters in the even/odd sectors are the dual longitudinal magnetizations  $\langle\tau_{x,y}\rangle$ or $\langle\mu_{x,y}\rangle$ which are accompanied by the spontaneous breaking of the $\mathbb{Z}_2$ symmetry of the corresponding chain Hamiltonian. The chain non-local string order $\mathcal{O}_{z,\tau/\mu}$ (known also as the parity order parameter \cite{Berg:2008,Montorsi:2012}) can also be linked via additional duality transformations to the $\mathbb{Z}_2 \otimes \mathbb{Z}_2$ symmetry breaking in superimposed Ising models \cite{Shelton:1996}. The string order $\mathcal{O}_{z,\tau/\mu}$ is called oscillating since it appears via non-decaying oscillations of the string-string correlation functions with the period of four dual lattice spacings \cite{Chitov:2019}.

The dual order parameters are worked out back to the order parameters of the original spin ladder. The staggered ladder possesses four phases with conventional antiferromagnetism ($m_{x,y} \neq 0$), three phases with the $z$-brane order, and two phases where the order parameters are $z$-branes with the pair of spins $\sigma^x,\sigma^y$ attached to the edge. Since we were able to find new exact results for the local and string order parameters in the XY chain with alternating field (see Appendix~\ref{AppB}), all the magnetizations and
the brane order parameters for the staggered ladder are found analytically as functions of the renormalized couplings of the effective mean-field Hamiltonian.

The ground-state phase diagram of the columnar ladder does not possess magnetic long-ranged order and demonstrates only the brane order. The order parameters of the three columnar phases are of the type presented in Table~\ref{Table1} and have their counterparts among the brane-ordered phases of the staggered ladder. Since the effective Hamiltonian of the columnar ladder maps onto two (equivalent) XY chains with alternating field and \textit{dimerization}, no analytical results are available for the order parameters in the latter case. The brane parameters of the columnar ladder are calculated numerically from the Toeplitz determinants.

All brane-ordered phases of the ladders with two dimerization patterns are spin liquids with distinct non-local order parameters, identified and calculated in this study. In particular, the phase (I) with the order parameter $\mathcal{O}_{z,1}$ detected for both types of dimerization
(see Figs.~\ref{PhaseDiag},\ref{Phasecol}) is the spin-liquid phase of the $SU(2)$-invariant homogeneous ladder (the case $\gamma=\delta=0$).

The calculation of topological numbers (winding number $N_w$ and the  number of the Majorana zero-energy edge modes $N_{\s M}$) is straightforward after the effective Hamiltonian is mapped onto the even and odd XY chains. It was done for all phases and for both dimerization patterns, and is particularly simple when evaluated via analytical continuation onto the complex wave numbers. $N_w$ or $N_{\s M}$ can change only when a root (or roots) of zero energy $z_\pm$ crosses the unit circle $|z|=1$ on the complex plane. The latter coincides with the condition for a quantum phase transition \cite{Verresen:2018,Chitov:2021}. In this sense the topological numbers  $N_w$ are  $N_{\s M}$ are complementary parameters, i.e., they do not provide additional information on criticality, which is not encoded in the complex roots of the model's spectra, or in its Lee-Yang zeros, when the temperature is finite \cite{Chitov:2021,Chitov:2022}.

The disorder lines (or modulation transitions) for both dimerization patterns are found as special points on the phase diagrams, where the complex conjugate roots $z_\pm$ in the even or odd sectors of the Hamiltonian merge and become degenerate. The points of disentanglement on the phase diagram where the ladder's ground state is factorized, are found on the intersections of the even and odd disorder lines.

An important conclusion of the present work is that the spin-Peierls ladders which are expected to dimerise into the energetically-favorable columnar pattern \cite{Chitov:2008}, can undergo a topological quantum phase transition ($\Delta N_{\s M} =\pm 4$) with gap closure, distinct brane orders (see Fig.~\ref{Phasecol}) when the microscopic  parameters (anisotropy, dimerization, rung vs leg couplings) are varied. We hope that our results will motivate the search for the real compounds which fit the profile.

%
\begin{acknowledgments}
G.Y.C. gratefully acknowledges financial support from Institut quantique (IQ) of
Universit\'{e} de Sherbrooke and Regroupement qu\'{e}b\'{e}cois sur les mat\'{e}riaux de pointe (RQMP).
\end{acknowledgments}

\begin{appendix}
\section{Mean-field theory: Technical details}\label{AppA}
%
%

We fermionize the spin ladder Hamiltonian \eqref{Ham} using the Jordan-Wigner  transformation (JWT). There are different ways to introduce this transformation when we depart from a single-chain problem to spin ladder (see, e.g., \cite{Azzouz:1993,*Azzouz:1994,Xiang:2007,NussinovChen:2008}. We use the snake-like JWT path
used in our earlier work \cite{Chitov:2017JSM} (see Fig.~\ref{SnakePath}) and proved to be convenient to deal with the two-leg ladder. Such JWT yields
the following fermionic Hamiltonian:
\begin{widetext}
\begin{eqnarray}
  H &=& \frac12 \sum_{n,\alpha} \Big\{e^{i\hat{\Phi}_{\alpha}(n)}\Big(J_{\alpha}(n)
    c^{\dag}_{\alpha}(n) c_{\alpha}(n+1)+\frac{J\gamma}{2}c^{\dag}_{\alpha}(n) c^{\dag}_{\alpha}(n+1)\Big)
    +2J_{\alpha}(n)\Big(\hat{n}_{\alpha}(n) -\frac12 \Big)\Big(\hat{n}_{\alpha} (n+1)-\frac12 \Big)\Big\} \nonumber \\
   &+& \frac12 \sum_{n} \Big\{ J_\perp(n) c^{\dag}_{1} (n)c_{2}(n)
   +2J_\perp(n)\Big(\hat{n}_{1}(n)-\frac12 \Big)\Big(\hat{n}_{2}(n) -\frac12 \Big)\Big\} + H.c.
\label{Ham2L2}
\end{eqnarray}
\end{widetext}
The phase operators $\hat{\Phi}_{\alpha}(n)$ appearing in Eq.~\ref{Ham2L2} are explicitly given in \cite{Chitov:2017JSM}.

The Hartee-Fock approximation for the fermionic interacting terms in \eqref{Ham2L2} is based on the general decoupling \cite{Santos:1989} for the product of two number operators ($\hat n_l=c_l^\dag c_l$):
\begin{eqnarray}
\label{HFA}
  \hat n_l \hat n_m &\approx&  \frac12 \hat n_l  + \frac12 \hat n_m   - \frac14  \nonumber \\
  &+& c_l^\dag c_m \langle  c_l c_m^\dag  \rangle   + \mathrm{H.c.} + |\langle  c_l c_m^\dag  \rangle |^2  \nonumber \\
  &+& c_l^\dag c_m^\dag  \langle  c_m c_l  \rangle   + \mathrm{H.c.} - |\langle  c_l c_m  \rangle |^2~.
\end{eqnarray}
The decoupled Hamiltonian reads
\begin{widetext}
\begin{eqnarray}
  H_{\s MF} &=& 2NC +\frac12\sum_{n,\alpha}\Big\{J_{\alpha}(n)\Big[
e^{i\hat{\Phi}_{\alpha}(n)}+2\langle c_{\alpha} (n) c_{\alpha}^{\dag}(n+1)\rangle\Big]
c^{\dag}_{\alpha}(n) c_{\alpha}(n+1) \nonumber  \\
  &+& J\gamma\Big[
e^{i\hat{\Phi}_{\alpha}(n)}-2\langle c_{\alpha} (n) c_{\alpha}(n+1)\rangle\Big]c^{\dag}_{\alpha}(n) c^{\dag}_{\alpha}(n+1)\Big\}
  + \frac12  J_{\perp}\sum_{n}[1+2t_{\perp}]c^{\dag}_{1}(n)c_{2}(n)+H.c.
\label{HFHam}
\end{eqnarray}
where
\begin{equation}
\label{tperpeq}
t_{\perp}(n)=\langle c_{1}(n)c^{\dag}_{2}(n)\rangle
=t_{\perp},
\end{equation}

According to the Lieb theorem \cite{Lieb:1994} the ground state of a quadratic fermionic Hamiltonian on a bipartite lattice at half-filling is the $\pi$-flux phase.
Imposing this requirement on the approximate Hamiltonian \eqref{HFHam} amounts to the approximation $e^{i\hat{\Phi}_{\alpha}(n)} \approx (-1)^{n+\alpha-1}$, thus
\begin{eqnarray}
  e^{i\hat{\Phi}_{\alpha}(n)}+2\langle c_{\alpha} (n) c_{\alpha}^{\dag}(n+1)\rangle &=&  (-1)^{n+\alpha-1} (1+2t_{\alpha}(n))~,\\
  e^{i\hat{\Phi}_{\alpha}(n)}-2\langle c_{\alpha} (n) c_{\alpha}(n+1)\rangle &=& (-1)^{n+\alpha-1} (1-2P_{\alpha}(n))~,
\end{eqnarray}
where we introduce the mean-field averaged parameters
\begin{equation}
\label{talphaeq}
t_{\alpha}(n)= \left\{
\begin{array}{lr}
\mathcal{K}+(-1)^{n+\alpha}\delta\eta,~~&~~~\mbox{(staggered)}\\
\mathcal{K}+(-1)^{n}\delta\eta,~~~&~~~\mbox{(columnar)},\\
\end{array}
\right.
\end{equation}
\begin{equation}
\label{Palphaeq}
P_{\alpha}(n)= \left\{
\begin{array}{lr}
P-(-1)^{n+\alpha}\delta\eta_p,~~&~~~\mbox{(staggered)}\\
P-(-1)^{n}\delta\eta_p,~~~&~~~\mbox{(columnar)}.\\
\end{array}
\right.
\end{equation}
The renormalized couplings \eqref{tR}-\eqref{raR} are to be found from the set of self-consistent equations obtained from minimization of the free-energy.
We present the equations for the case $\gamma_a=\gamma_{a\s R}=0$.

The sets of the mean-field equations are different for two dimerization patterns.\\

\underline{\emph{(1) Staggered dimerization:}}\\

Minimization of the ground-state energy evaluated with the spectrum \eqref{spectra_s} yields the bond average
\begin{eqnarray}
\label{KMF}
  \mathcal{K} = \frac{ t_{\s R}}{4\pi}\int_{0}^{\pi/2} dk \cos^2k \Big\{
   \frac{1}{E_{++}}+ \frac{1}{E_{+-}}+ \frac{1}{E_{-+}}+\frac{1}{E_{--}} \Big\}
\end{eqnarray}
and the dimerization susceptibility $\eta$
\begin{eqnarray}
\label{etaMF}
 \delta  \eta = \frac{1}{4\pi}\int_{0}^{\pi/2} dk \sin^2k &\phantom& \Big\{\delta_{\s R}
 \Big(\frac{1}{E_{++}}+ \frac{1}{E_{+-}}+ \frac{1}{E_{-+}}+\frac{1}{E_{--}} \Big)
 +\gamma_{\s R}
 \Big(\frac{1}{E_{++}}+ \frac{1}{E_{+-}}- \frac{1}{E_{-+}}-\frac{1}{E_{--}} \Big) \nonumber \\
 &+& \frac{J_{\perp \s R}}{2 \sin k}
 \Big(\frac{1}{E_{++}}- \frac{1}{E_{+-}}- \frac{1}{E_{-+}}+\frac{1}{E_{--}} \Big) \Big\} ~.
\end{eqnarray}
The anomalous pairing amplitude is found as
\begin{eqnarray}
\label{PMF}
 P =  \frac{1}{4\pi}\int_{0}^{\pi/2} dk \sin^2k &\phantom& \Big\{\gamma_{\s R}
 \Big(\frac{1}{E_{++}}+ \frac{1}{E_{+-}}+ \frac{1}{E_{-+}}+\frac{1}{E_{--}} \Big)
 +\delta_{\s R}
 \Big(\frac{1}{E_{++}}+ \frac{1}{E_{+-}}- \frac{1}{E_{-+}}-\frac{1}{E_{--}} \Big) \nonumber \\
 &+& \frac{J_{\perp \s R}}{2 \sin k}
 \Big(\frac{1}{E_{++}}- \frac{1}{E_{+-}}+ \frac{1}{E_{-+}}-\frac{1}{E_{--}} \Big) \Big\} ~,
\end{eqnarray}
and the transverse bond parameter is:
\begin{eqnarray}
\label{tperpMF}
   t_\perp =  \frac{1}{4\pi}\int_{0}^{\pi/2} dk &\phantom& \Big\{
   \frac12 J_{\perp \s R} \Big(\frac{1}{E_{++}}+ \frac{1}{E_{+-}}+ \frac{1}{E_{-+}}+\frac{1}{E_{--}} \Big)
  +\delta_{\s R} \sin k
 \Big(\frac{1}{E_{++}}- \frac{1}{E_{+-}}- \frac{1}{E_{-+}}+\frac{1}{E_{--}} \Big) \nonumber \\
 &+& \gamma_{\s R} \sin k
 \Big(\frac{1}{E_{++}}- \frac{1}{E_{+-}}+ \frac{1}{E_{-+}}-\frac{1}{E_{--}} \Big) \Big\} ~.
\end{eqnarray}
~\\

\underline{\emph{(2) Columnar dimerization:}}\\

Minimization of the ground-state energy with the spectrum \eqref{spectra_c} yields the following set of the mean-field equations:

\begin{eqnarray}
\label{KMFc}
  \mathcal{K} &=& \frac{ t_{\s R}}{2\pi}\int_{0}^{\pi/2} dk \cos^2k
  \Big\{ \frac{1}{E_{+}}+ \frac{1}{E_{-}} \Big\}~, \\
\label{etaMFc}
 \delta  \eta &=& \frac{\delta_{\s R}}{2\pi}\int_{0}^{\pi/2} dk \sin^2k  \Big\{
 \Big(\frac{1}{E_{+}}+ \frac{1}{E_{-}} \Big) +
 \frac{\gamma_{\s R} \sin k}{R}
 \Big(\frac{1}{E_{+}}- \frac{1}{E_{-}} \Big) \Big\} ~,\\
 \label{PMFc}
 P &=&  \frac{1}{2\pi}\int_{0}^{\pi/2} dk \sin^2 k  \Big\{
 \gamma_{\s R}  \Big(\frac{1}{E_{+}}+ \frac{1}{E_{-}} \Big) +
 \frac{R}{\sin k} \Big(\frac{1}{E_{+}}- \frac{1}{E_{-}} \Big) \Big\} ~,\\
 \label{tperpMFc}
 t_\perp &=& \frac{J_{\perp \s R}}{4\pi}\int_{0}^{\pi/2} dk   \Big\{
 \Big(\frac{1}{E_{+}}+ \frac{1}{E_{-}} \Big) +
 \frac{\gamma_{\s R} \sin k}{R}
 \Big(\frac{1}{E_{+}}- \frac{1}{E_{-}} \Big) \Big\} ~.\\
\end{eqnarray}
\end{widetext}
where $R \equiv \sqrt{\delta_{\s R}^2 \sin^2 k+ \frac14 J_{\perp \s R}^2}$.
In the limit $J_\perp=0$ the above equations agree with those we found for the XYZ chain \cite{Chitov:2020}.

\begin{figure}[h]
\includegraphics[width=8.0 cm]{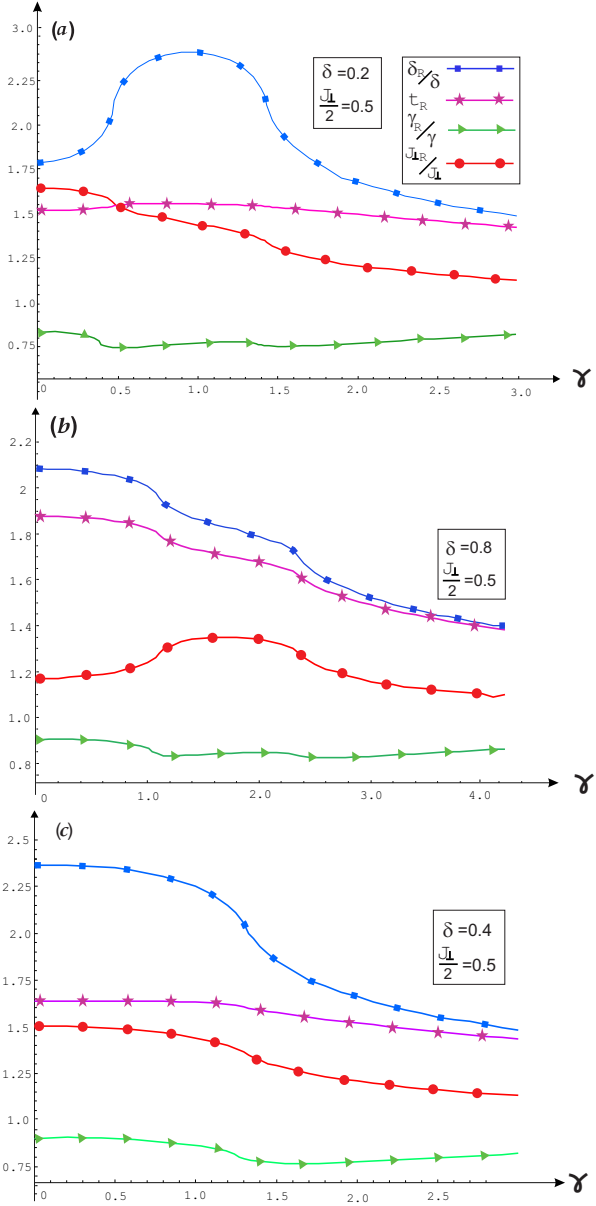}
\caption{The ratios of the renormalized couplings with respect to their bare values calculated
from the mean-field equations. The parameters are calculated along two paths of constant $\delta$ for the staggered ladder (a,b)
and one path of constant $\delta$ for the columnar ladder (c). The paths resemble those shown in Figs.~\ref{PhaseDiag} and \ref{Phasecol},
but taken on the plane of bare parameters ($\delta,\gamma$).}
\label{RCouples}
\end{figure}

From numerical solution of the above mean-field equations and evaluation of the renormalized couplings \eqref{tR}-\eqref{rR},
we find that the renormalized and the bare values differ by a factor of order of unity. It is important to distinguish these two sets of couplings for making quantitative comparisons of, e.g., phase boundaries, gap values, etc,  with the numerical simulations. However to study the qualitative physical properties of the phases, transitions between them, the local and non-local order parameters, it is convenient to present results directly in terms of the renormalized parameters, as it is done mostly in this paper. The ratios of the renormalized and bare parameters of the model calculated for several cases are presented in Fig.~\ref{RCouples}.

%
%
%
%

\section{Duality of two transverse-field XY chain models and order parameters }\label{AppB}
%
%

In this Appendix we will present the canonical transformations to map the spin-$\frac12$ XY chain with a uniform transverse field onto
the chain in a staggered field. This mapping is used to get analytical results for the order parameters.

First, we find explicit expressions for the longitudinal magnetizations $m_{x,y}$ in the chain with a staggered field. It is an interesting and useful result, since a brut force calculation of such parameters in a chain with two lattice-space periodicity, amounts to evaluation of limiting values of the determinants of $2 \times 2$-block Toeplitz matrices \cite{Chitov:2019}. The latter problem seems to be quite hopeless for analytical calculations \cite{Widom:1976,Its:2013,Basor:2019}, but it is well amenable via mappings onto the chain with a uniform field, as shown below.

Second, we calculate for the first time the string order parameter in a closed form for the XY chain in a uniform transverse field. Using the mapping we get the explicit formulas for the oscillating string order parameter in the topological phase of the staggered model, bypassing a problem of block Toeplitz matrices.

Since the effective mean-field Hamiltonian of the two-leg ladder maps onto a sum of two XY chains in the alternating field,
as shown in the main text of the paper, the results of this Appendix allow us to find analytic expressions for the brane order parameters of the staggered ladder.

 The Hamiltonian of the XY chain in uniform transverse field is defined as:
 \begin{equation}
 H=\sum_{n=1}^{N}\frac{J}{4}\Big\{(1+\gamma)\sigma_{n}^{x}\sigma_{n+1}^{x}
 +(1-\gamma)\sigma_{n}^{y}\sigma_{n+1}^{y}\Big\}+\frac{h}{2}\sigma_{n}^{z}
 \label{Hamxy}
 \end{equation}
 In this Appendix we return to the use of dimensionful units and restore the exchange coupling $J$.
 The properties of this model (\ref{Hamxy}) are well known  \cite{McCoyII:1971,Franchini:2017}. Its spectrum is
 \begin{equation}
 E(k)=J\sqrt{\Big(\frac{h}{J}-\cos k\Big)^2+\gamma^{2}\sin^{2} k}
 \label{Ek}
 \end{equation}
with $k \in [-\pi, \pi]$. In the range $|h/J|<1$ the model is antiferromagnetically ($J>0$) ordered :
$\langle\sigma_{L}^{x}\sigma_{L+n}^{x}\rangle \rightarrow \pm m_{x}^{2}$ as $n \rightarrow \infty$, with the spontaneous longitudinal
magnetization
\begin{equation}
m_{x}^{2}=\frac{2}{1+\gamma}\Big\{\gamma^{2}\Big(1-\Big(\frac{h}{J}\Big)^{2}
\Big)\Big\}^{1/4}~,
\label{mx}
\end{equation}
when $\gamma>0$. At $\gamma <0$ the order changes: $m_{x} \leftrightarrow m_{y}$. In the polarized phase $h/J>1$ (the phase diagram is symmetric with respect to the sign change of the field) only induced magnetization $m_z$ is present. However one can notice \cite{Chitov:2019,Chitov:2020} appearance of a continuous
monotonous $z$-string order parameter $\mathcal{O}_z$ defined from the limit of the string-string  correlator
\begin{equation}
\mathfrak{D}_{zz}(L,R) \equiv \Big\langle\prod_{n=L}^{R} \sigma_{n}^{z}\Big\rangle
\xrightarrow[\s R-L \to \infty]{~} \mathcal{O}_z^2
\label{SOPz}
\end{equation}
To find $\mathcal{O}_z$ we apply the standard techniques to calculate Toeplitz determinants using the Szeg\"{o}'s theorem
\cite{McCoy:2010}. The result is expressed via two roots $\lambda_\pm$ of the model's spectra on the complex momentum plane:
\begin{equation}
\label{OzAnG}
  \mathcal{O}_z^2= \Big[ \frac{(1-\lambda_-^2)(\lambda_+^2-1)}{(\lambda_+-\lambda_-)^2} \Big]^{1/4}~,
\end{equation}
where
\begin{equation}
\label{lampm}
\lambda_\pm = \frac{h/J \pm \sqrt{(h/J)^2+\gamma^2-1}}{1+\gamma}~,
\end{equation}
yielding
\begin{equation}
\label{OzAn}
  \mathcal{O}_z^2= \Big[ \frac{(h/J)^2-1}{(h/J)^2+\gamma^2-1} \Big]^{1/4}~.
\end{equation}
This $z$-string parameter, known also (up to irrelevant prefactors) as the parity string order parameter  \cite{Berg:2008,Montorsi:2012,Rath:2013}, vanishes at the boundary of the polarized phase with the correct critical index of the order parameter, i.e. $\mathcal{O}_z \propto (h/J-1)^{1/8}$. In the isotropic limit $\gamma=0$ it becomes a plateau $\mathcal{O}_z=1$ with a discontinuity of the phase boundary \cite{Chitov:2019,Chitov:2020}, similar to the plateau of induced magnetization $m_z$.

With the canonical transformations \eqref{CanX}-\eqref{CanZ} where $\tau \Leftrightarrow \sigma $,
we can map  the chain (\ref{Hamxy}) onto the XY chain with staggered magnetic field:
\begin{equation}
 H=\sum_{n=1}^{N}\frac{\tilde{J}}{4}\Big\{(1+\tilde{\gamma})\tilde{\sigma}
 _{n}^{x}\tilde{\sigma}_{n+1}^{x}+(1-\tilde{\gamma})\tilde{\sigma}_{n}^{y}
 \tilde{\sigma}_{n+1}^{y}\Big\}+\frac{1}{2}(-1)^{n}\tilde{h}_{a}\tilde{\sigma}_{n}^{z}~,
 \label{Hamgha}
 \end{equation}
where the corresponding parameters of two Hamiltonians are related as:
\begin{eqnarray}
\label{Correspondence}
  \tilde{J} &=& \gamma J \nonumber \\
   \tilde{\gamma} &=& \frac{1}{\gamma} \\
  \tilde{h}_{a} &=& h~. \nonumber
\end{eqnarray}
The dimerized XY chain with uniform and staggered fields was studied recently \cite{Chitov:2019} in great detail, so we can easily recover some key properties
of the model (\ref{Hamgha}) as a simpler special case. The spectrum of the fermionized Hamiltonian (\ref{Hamgha}) has two eigenvalues
\begin{equation}
 \tilde{E}_\pm (k)=\tilde{J}\sqrt{\Big(\frac{\tilde{h}_a}{\tilde{J}}\pm \tilde{\gamma}\sin k\Big)^2+\cos^{2} k}
 \label{Epmha}
\end{equation}
where the wavenumbers lie in the reduced Brillouin zone $k \in [-\pi/2, \pi/2]$. Using the correspondence \eqref{Correspondence},
wavenumber shifts by $\pm \pi/2$, and the Brullouin zone unfolding, one can establish the equivalence between the spectra \eqref{Ek} and \eqref{Epmha}, as well as  an equal number of eigenmodes in two spectra.

The model (\ref{Hamgha}) has two different phases:

1. At $\tilde h_a/\tilde J < \tilde \gamma$ it has a local magnetic order $\tilde m_x$. Since $\sigma_n^x =\tilde \sigma_n^x$ and according to \eqref{Correspondence}: $\tilde h_a/\tilde J < \tilde \gamma~$
$\longleftrightarrow h/J<1$, the phase $\tilde m_x \neq 0$ of (\ref{Hamgha}) corresponds to the phase $m_x \neq 0$ of \eqref{Hamxy}. From \eqref{mx} and \eqref{Correspondence} we can calculate the magnetic order parameter in the staggered model:
\begin{equation}
\tilde m_{x}^{2}=\frac{2}{1+\tilde \gamma}
\Big\{\tilde \gamma^{2}-\big(\tilde h_a /\tilde J \big)^{2}\Big\}^{1/4}
~~ \mathrm{at}~~\tilde \gamma > \tilde h_a/\tilde J~.
\label{mxAlt}
\end{equation}
The magnetization $\tilde m_y$ at $\tilde \gamma < - \tilde h_a/\tilde J$ can be calculated from the symmetry $\tilde m_y(-\tilde \gamma) =\tilde m_x(\tilde \gamma)$.

2. In the region  $-\tilde h_a/ \tilde J < \tilde \gamma <\tilde h_a/\tilde J$ the magnetization disappears, and non-local order detected by non-vanishing oscillations of the string-string correlation function $\tilde{\mathfrak{D}}_{zz}$ with a period of four lattice spacings is found from numerical calculations \cite{Chitov:2019}. This oscillating string order is parameterized as follows
\cite{Chitov:2019}:
\begin{widetext}
\begin{equation}
\label{Oz123}
 \tilde{\mathfrak{D}}_{zz}(L,R)
 \xrightarrow[R \to \infty]{~}
  \left\{
    \begin{array}{c}
      (-1)^m \mathcal{O}_{z,1}^2~,~~~~~ L=1,~R=2m \\[0.2cm]
      (-1)^m \mathcal{O}_{z,2}^2~,~~~~ L=2,~R=2m \\[0.2cm]
      (-1)^{m+L} \mathcal{O}_{z,3}^2~,~~~~~ L=1,~R=2m+1~ \mathrm{or}~ L=2,~R=2m+1  \\
    \end{array}
   \right.
\end{equation}
\end{widetext}
In a special case of zero dimerization as the Hamiltonian \eqref{Hamgha} under consideration, the string correlator oscillates with a constant amplitude $\mathcal{O}_{z,1}=\mathcal{O}_{z,2}=\mathcal{O}_{z,3}$ following the pattern $(++--)$ \cite{Chitov:2019}.

One can easily check that the region with the oscillating string order of the staggered model \eqref{Hamgha} corresponds to the polarized phase of the uniform chain (\ref{Hamxy}) at $|h/J|>1$. The uniform string order parameter \eqref{SOPz} in that case is dual to the oscillating string parameters of
the staggered model. Indeed, since $\tilde \sigma_n^z = (-1)^n \sigma_n^z$ we can easily establish relation between string correlators in two models:
\begin{equation}
\label{DzDz}
  \tilde{\mathfrak{D}}_{zz}(L,R) =(-1)^s \mathfrak{D}_{zz}(L,R)~,
\end{equation}
where $s \equiv (R+L)(R-L+1)/2$. Taking $L=1$ we find the oscillating string order as:
\begin{equation}
\label{Oz}
 \tilde{\mathfrak{D}}_{zz}(1,R)
 \xrightarrow[\s R \to \infty]{~}
  \left\{
    \begin{array}{c}
      (-1)^m \mathcal{O}_{z}^2~,~~R=2m \\[0.2cm]
      (-1)^{m+1} \mathcal{O}_{z}^2~,~~R=2m+1 \\
    \end{array}
   \right.
\end{equation}
with
\begin{equation}
\label{OzAnSt}
  \mathcal{O}_z^2= \Big[ \frac{(\tilde{h}_a/ \tilde{J})^2-\tilde{\gamma}^2}{(\tilde{h}_a/\tilde{J})^2-\tilde{\gamma}^2+1} \Big]^{1/4}~ \mathrm{at}~ -\tilde h_a/\tilde J<\tilde \gamma < \tilde h_a/\tilde J~.
\end{equation}

\end{appendix}
\bibliography{C:/Papers/BibRef/CondMattRefs}
%
%
%
%
\end{document}